\documentclass[english]{article}
\usepackage[T1]{fontenc}
\usepackage[latin9]{inputenc}
\usepackage[a4paper]{geometry}
\usepackage{amsmath}
\usepackage{amssymb}
\usepackage{graphicx}
\usepackage{setspace}
\usepackage{esint}
\makeatletter

\providecommand{\tabularnewline}{\\}

\numberwithin{equation}{section}

\usepackage{cite}

\usepackage{babel}

\makeatother

\usepackage{babel}
\begin{document}
\begin{titlepage}

\global\long\def\thefootnote{\fnsymbol{footnote}}

\begin{flushright}
\begin{tabular}{l}
UTHEP-692 \tabularnewline
\end{tabular}
\par\end{flushright}

\bigskip{}

\begin{center}
\textbf{\Large{}{}{}{}{}{}String field theory solution corresponding
to constant background magnetic field}{\Large{}{} {}{}{}} 
\par\end{center}

\bigskip{}

\begin{center}
{\large{}{}{}{}{}{}{}{}{}}Nobuyuki Ishibashi\footnote{e-mail: ishibash@het.ph.tsukuba.ac.jp},
Isao Kishimoto\footnote{e-mail: ikishimo@ed.niigata-u.ac.jp}, Tomohiko
Takahashi\footnote{e-mail: tomo@asuka.phys.nara-wu.ac.jp} {\large{}{}{}{}{}{}{}{}{}} 
\par\end{center}

\begin{center}
Center for Integrated Research in Fundamental Science and Engineering
(CiRfSE),\\
 Faculty of Pure and Applied Sciences, University of Tsukuba\\
 Tsukuba, Ibaraki 305-8571, JAPAN\\
 Faculty of Education, Niigata University, Niigata 950-2181, JAPAN\\
 Department of Physics, Nara Women's University, Nara 630-8506, JAPAN 
\par\end{center}

\bigskip{}

\bigskip{}

\bigskip{}

\begin{abstract}
Following the method recently proposed by Erler and Maccaferri, we
construct solutions to the equation of motion of Witten's cubic string
field theory, which describe a constant magnetic field background. We
study the boundary condition changing operators relevant to such a background,
and calculate their operator product expansions. We obtain solutions
whose classical action coincide with the Born-Infeld action. 
\end{abstract}
\global\long\def\thefootnote{\arabic{footnote}}

\end{titlepage}

\section{Introduction}

Since the discovery of the tachyon vacuum solution \cite{Schnabl2006},
the classical solutions of cubic string field theory \cite{Witten1986}
have been studied quite actively \cite{Ellwood2007,Erler2009,Kiermaier2008,Schnabl2007,Fuchs2007,Kiermaier2009a,Erler2007b,Okawa2007a,Okawa2007,Fuchs2008,Fuchs2007a,Kiermaier2009,Kiermaier2008b,Kiermaier2011,Takahashi2002,Takahashi2001,Kishimoto2002a,Bonora2011d,Bonora2011b}.
In a recent paper, Erler and Maccaferri \cite{Erler2014} have given
a way to realize any boundary conformal field theory (BCFT) as a solution
of the equation of motion.

With the method of Erler and Maccaferri, it is now possible to study
various open string backgrounds from the point of view of string field
theory. It has been known for many years that the open string theory in
a constant magnetic field background \cite{Abouelsaood1987} has various
interesting features. It corresponds to an exactly solvable BCFT and
the action is given by the Born-Infeld action. Noncommutative geometry
appears in the open string theory around the background \cite{Connes1998,Seiberg1999}.

What we would like to do in this paper is to construct the Erler-Maccaferri
solution corresponding to the constant magnetic background. We start
from the string field theory associated with a D-brane with vanishing
gauge field background and would like to realize the constant magnetic
background as a solution to the equation of motion. In order to do
so, we study the boundary condition changing (BCC) operators which
change the open string boundary condition from that of one background
to another.\footnote{In a different context, it was shown that the Dirac-Born-Infeld factor
appears from the tachyon vacuum solution in a constant $B_{\mu\nu}$
background \cite{Ishida:2008jc}. There is also an earlier related
work \cite{deGroot:1989in}, which seems rather formal. }

The organization of this paper is as follows. In section \ref{sec:Erler-Maccaferri-solutions},
we briefly review the Erler-Maccaferri solutions. In section \ref{sec:BCC-operators-for},
we study the BCC operators necessary for the construction of our solution.
We investigate the open string states corresponding to these operators,
calculate their correlation functions, and finally obtain their
operator product expansions. In section \ref{sec:Classical-solutions-for},
we construct the solutions of the equation of motion of string field theory
using the BCC operators studied in section \ref{sec:BCC-operators-for}.
Section \ref{sec:Concluding-remarks} is devoted to conclusions and
discussions. In the appendices, details of calculations concerning
the BCC operators are exhibited.

\section{Erler-Maccaferri solutions\label{sec:Erler-Maccaferri-solutions}}

\begin{figure}
\begin{centering}
\includegraphics[scale=0.8]{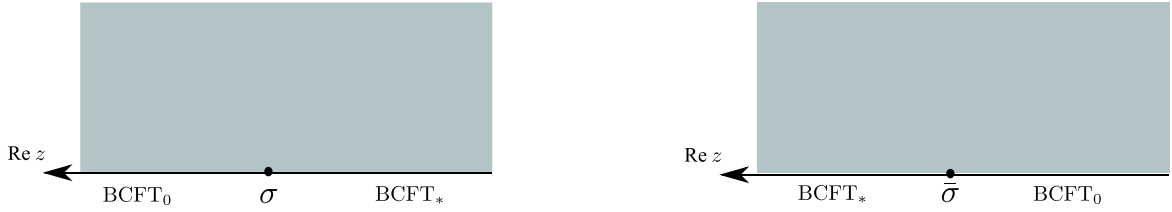} 
\par\end{centering}
\caption{$\sigma(z),\,\bar{\sigma}(z)$. The boundary on the left of $\sigma$
on the real axis corresponds to the boundary condition of $\mbox{BCFT}_{0}$
and the boundary on the right corresponds to that of $\mbox{BCFT}_{*}$,
and vice versa for $\bar{\sigma}$. The order of the star product and
that in the figure coincide if we take the direction of the real axis
as above \cite{Erler2009}. \label{fig:sigmasigmabar}}
\end{figure}

Let us first explain how to construct the Erler-Maccaferri solutions
to the equation of motion 
\begin{equation}
Q\Psi+\Psi^{2}=0\label{eq:eqm}
\end{equation}
of Witten's cubic string field theory. Here let us assume that the
solution $\Psi=0$ corresponds to a boundary conformal field theory
$\mbox{BCFT}_{0}$. We would like to discuss a solution associated
with another boundary conformal field theory $\mbox{BCFT}_{*}.$ The
solution is constructed from the string fields 
\[
K,B,c,\sigma,\bar{\sigma}\ .
\]
The fields $K,B,c$ are the ones which play crucial roles in the construction
of classical solutions of cubic string field theory \cite{Schnabl2006,Okawa2006}
and satisfy the so-called $KBc$ algebra: 
\begin{eqnarray}
B^{2}=c^{2}=0\,, & \left[K,B\right]=0\,, & Bc+cB=1\,,\nonumber \\
QK=0\,, & QB=K\,, & Qc=c\partial c\,.\label{eq:KBc}
\end{eqnarray}
Here the product of the string fields is the star product and $Q\left(\cdot\right)$
denotes the BRST variation. The string fields $\sigma,\bar{\sigma}$
are the ones which can be expressed by 
\begin{eqnarray*}
\sigma & = & \sigma(1)I\,,\\
\bar{\sigma} & = & \bar{\sigma}(1)I\,,
\end{eqnarray*}
where $I$ is the identity string field and $\sigma(s),\bar{\sigma}(s)$
are the BCC operators such that $\sigma(s)$ changes the open string
boundary condition from the one corresponding to $\mbox{BCFT}_{*}$
to the one corresponding to $\mbox{BCFT}_{0}$, and $\bar{\sigma}(s)$
changes in reverse, as indicated in figure \ref{fig:sigmasigmabar}.
We assume that $\sigma(s),\,\bar{\sigma}(s)$ are matter primary fields
of weight $0$ and satisfy the operator product expansion (OPE) 
\begin{eqnarray}
\bar{\sigma}(s)\sigma(0) & \sim & 1\,,\nonumber \\
\sigma(s)\bar{\sigma}(0) & \sim & \frac{g_{*}}{g_{0}}\label{eq:OPEsigma}
\end{eqnarray}
for $s>0$, where $g_{*}$ is the disk partition function of $\mbox{BCFT}_{*}$
and $g_{0}$ is that of $\mbox{BCFT}_{0}$. We have the following algebraic
relations involving $\sigma,\,\bar{\sigma}$\,:
\begin{eqnarray}
\left[\sigma,c\right]=0\,, & \left[\sigma,\partial c\right]=0\,, & \left[\sigma,B\right]=0\,,\nonumber \\
\left[\bar{\sigma},c\right]=0\,, & \left[\bar{\sigma},\partial c\right]=0\,, & \left[\bar{\sigma},B\right]=0\,,\nonumber \\
Q\sigma=c\partial\sigma\,, & Q\bar{\sigma}=c\partial\bar{\sigma}\,,\nonumber \\
\bar{\sigma}\sigma=1\,, & \sigma\bar{\sigma}=\frac{g_{*}}{g_{0}}\,.\label{eq:sigmasigmabar}
\end{eqnarray}

With all these string fields, it is possible to construct a solution
$\Psi$ given by 
\begin{equation}
\Psi=\Psi_{\mathrm{tv}}-\Sigma\Psi_{\mathrm{tv}}\bar{\Sigma}\,,\label{eq:EelerMaccaferri}
\end{equation}
where $\Psi_{\mathrm{tv}}$ is the Erler-Schnabl solution \cite{Erler2009}
for the tachyon vacuum 
\[
\Psi_{\mathrm{tv}}=\frac{1}{\sqrt{1+K}}c(1+K)Bc\frac{1}{\sqrt{1+K}}\,,
\]
and $\Sigma,\,\bar{\Sigma}$ are the string fields 
\begin{eqnarray}
\Sigma & = & Q_{\Psi_{\mathrm{tv}}}\left(\frac{1}{\sqrt{1+K}}B\sigma\frac{1}{\sqrt{1+K}}\right)\,,\nonumber \\
\bar{\Sigma} & = & Q_{\Psi_{\mathrm{tv}}}\left(\frac{1}{\sqrt{1+K}}B\bar{\sigma}\frac{1}{\sqrt{1+K}}\right)
\label{eq:Sigma}
\end{eqnarray}
with 
\[
Q_{\Psi_{\mathrm{tv}}}=Q+\left[\Psi_{\mathrm{tv}},\ \right]\,.
\]
Using the formula 
\[
\bar{\Sigma}\Sigma=1\,,
\]
which can be derived from (\ref{eq:KBc}), (\ref{eq:sigmasigmabar}), and
(\ref{eq:Sigma}), it is possible to calculate the energy and the
gauge-invariant observable \cite{Ellwood2008,Kawano:2008ry,Kawano:2008jv,Baba:2012cs,Kudrna:2012re}
as 
\begin{eqnarray*}
 &  & E=\frac{1}{\mathrm{Vol}(X^{0})}\left(-\frac{g_{0}}{2\pi^{2}}+\frac{g_{*}}{2\pi^{2}}\right)\,,\\
 &  & \mathrm{Tr}_{\mathcal{V}}\left[\Psi\right]=\frac{1}{4\pi i}\left(\left\langle \mathcal{V}\right|c_{0}^{-}\left|B_{0}\right\rangle -\left\langle \mathcal{V}\right|c_{0}^{-}\left|B_{*}\right\rangle \right)\,.
\end{eqnarray*}
These results imply that the solution corresponds to the boundary
conformal field theory $\mbox{BCFT}_{*}$.

Therefore, in order to construct a solution corresponding to $\mbox{BCFT}_{*}$,
we need to construct primary fields $\sigma(s),\,\bar{\sigma}(s)$
of weight $0$ satisfying the OPE (\ref{eq:OPEsigma}). Erler and
Maccaferri considered the case of time-independent solutions and $\sigma(s),\,\bar{\sigma}(s)$
are expressed as 
\[
\sigma(s)=\sigma_{*}e^{i\sqrt{h}X^{0}}(s),\ \bar{\sigma}(s)=\bar{\sigma}_{*}e^{-i\sqrt{h}X^{0}}(s)\,.
\]
Here $\sigma_{*}(s),\,\bar{\sigma}_{*}(s)$ are the BCC operators
of weight $h$ which satisfy 
\begin{eqnarray}
\bar{\sigma}_{*}(s)\sigma_{*}(0) & \sim & s^{-2h}\,,\nonumber \\
\sigma_{*}(s)\bar{\sigma}_{*}(0) & \sim & \frac{g_{*}}{g_{0}}s^{-2h}\label{eq:sigmastar}
\end{eqnarray}
for $s>0$, and they act as the identity operator in the time direction.

\section{BCC operators for a constant background magnetic field\label{sec:BCC-operators-for}}

We would like to construct a solution to the equation of motion (\ref{eq:eqm})
which corresponds to a constant magnetic field background. Here we
consider the bosonic string theory in $26$-dimensional Minkowski
space-time and take $\mbox{BCFT}_{0}$ to be the usual one for a D$p$-brane,
i.e. Neumann boundary conditions for $X^{\mu}\ (\mu=0,\cdots,p)$
and Dirichlet boundary conditions for $X^{I}\ (I=p+1,\cdots,25)$.
We take $\mbox{BCFT}_{*}$ to be the one with a constant magnetic
field, namely 
\begin{equation}
F_{\mu\nu}=\mathrm{constant}\,,\label{eq:constantmagnetic}
\end{equation}
with $F_{0\mu}=0$. Our goal is to study the BCC operators in this
setup and calculate the OPEs, which are necessary for the construction
of the Erler-Maccaferri solution.

\subsection{Canonical quantization\label{subsec:First-quantization section}}

The BCC operators correspond to states of open strings with one end
satisfying the $\mbox{BCFT}_{0}$ boundary condition and the other end
satisfying the $\mbox{BCFT}_{*}$ boundary condition. Therefore it is
possible to deduce properties of the BCC operators by studying the
open string states in such sectors. Let us consider the worldsheet
theory of the open string which is given by the conformal field theory
(CFT) on the strip where the Neumann boundary conditions are imposed
at $\sigma=0$ and the other boundary at $\sigma=\pi$ is coupled
to the electromagnetic fields as

\begin{equation}
\int A_{\mu}(X)\,\dot{X}^{\mu}|_{\sigma=\pi}\,dt.
\end{equation}
In the case at hand, we have a constant magnetic field (\ref{eq:constantmagnetic})
and we take 
\begin{equation}
A_{\mu}=-\frac{1}{2}F_{\mu\nu}X^{\nu}.\label{eq:gauge_fix}
\end{equation}
Since $F_{\mu\nu}$ is a real antisymmetric tensor with $F_{0\mu}=0$,
we can make a rotation to put it into block diagonal form 
\begin{equation}
F_{\mu\nu}=\left(\begin{array}{cccccc}
0\\
 & 0 & F_{12}\\
 & -F_{12} & 0\\
 &  &  & 0 & F_{34}\\
 &  &  & -F_{34} & 0\\
 &  &  &  &  & \ddots
\end{array}\right)\,.\label{eq:block}
\end{equation}
We concentrate on one of the blocks,
\[
\left(\begin{array}{cc}
0 & F_{12}\\
-F_{12} & 0
\end{array}\right)\,,
\]
and introduce complex coordinates, $X=(X^{1}+i\,X^{2})/\sqrt{2}$
and $\tilde{X}=(X^{1}-i\,X^{2})/\sqrt{2}$, in these two dimensions.
The boundary conditions for these variables turn out to be 
\begin{align}
X' & =0,\ \ \ \tilde{X}'=0\ \ \ \ \ \ \ \ \ \ \ \ \ \ (\,\sigma=0\,),\label{eq:bc1}\\
X' & =-i\,2\pi\alpha'F_{12}\,\dot{X},\ \ \ \tilde{X}'=i\,2\pi\alpha'F_{12}\tilde{X}\ \ \ (\,\sigma=\pi\,)\,.\label{eq:bc2}
\end{align}

The open string theory with such boundary conditions is studied in
\cite{Abouelsaood1987}, the results of which are reviewed in appendix
\ref{sec:First-quantization-of}. The mode expansions of $X$ and
$\tilde{X}$ are given by 
\begin{align}
X(z,\,\bar{z}) & =x+i\sqrt{\frac{\alpha'}{2}}\sum_{k=-\infty}^{\infty}\frac{1}{k+\lambda}(z^{-k-\lambda}+\bar{z}^{-k-\lambda})\,\alpha_{k+\lambda}\,,\label{eq:mode_a}\\
\tilde{X}(z,\,\bar{z}) & =\tilde{x}+i\sqrt{\frac{\alpha'}{2}}\sum_{k=-\infty}^{\infty}\frac{1}{k-\lambda}(z^{-k+\lambda}+\bar{z}^{-k+\lambda})\,\tilde{\alpha}_{k-\lambda}\,,\label{eq:mode}
\end{align}
where $z=e^{\tau+i\sigma}$ and $\lambda$ is related to the magnetic
field as $2\pi\alpha'F_{12}=\tan\pi\lambda\ (0\leq\lambda<1)$. In
order for $X^{1},X^{2}$ to be real, $(\alpha_{k+\lambda})^{\dagger}=\tilde{\alpha}_{-k-\lambda}$
and $\left(x\right)^{\dagger}=\tilde{x}$. $\alpha_{k+\lambda},\,\tilde{\alpha}_{k-\lambda},\,x$, and $\tilde{x}$
satisfy the commutation relations 
\begin{eqnarray}
[\alpha_{k+\lambda},\,\tilde{\alpha}_{k^{\prime}-\lambda}] & = & (k+\lambda)\delta_{k+k^{\prime},0}\,,\nonumber \\{}
[\alpha_{k+\lambda,}\,\alpha_{k^{\prime}+\lambda}] & = & [\tilde{\alpha}_{k-\lambda},\,\tilde{\alpha}_{k^{\prime}-\lambda}]=0\,,\label{eq:comrel}\\{}
[x,\,\tilde{x}] & = & -\frac{1}{F_{12}}=-\frac{2\pi\alpha'}{\tan\pi\lambda}\,.\label{eq:zero_comrel}
\end{eqnarray}
Equation (\ref{eq:zero_comrel}) implies that a noncommutative space arises
due to the background gauge field.

The energy-momentum tensor is given by 
\begin{equation}
T(z)=\lim_{z'\rightarrow z}\left[-\frac{2}{\alpha'}\,\partial X(z)\partial\tilde{X}(z')-\frac{1}{(z-z')^{2}}\right]\,,\label{eq:EM}
\end{equation}
and we get the Virasoro generators 
\begin{equation}
L_{n}=\sum_{k=-\infty}^{\infty}:\alpha_{k+\lambda}\tilde{\alpha}_{n-k-\lambda}:+\frac{1}{2}\lambda(1-\lambda)\delta_{n,0}\,,\label{eq:Virasoro}
\end{equation}
where $:\cdot:$ denotes the normal ordering. We construct a basis
of Fock space using the oscillators $\alpha_{k+\lambda},\,\tilde{\alpha}_{k-\lambda}$.

To every ket state, there corresponds a BCC operator which may be
used to construct the Erler-Maccaferri solution. In this paper, for
simplicity, we restrict ourselves to the BCC operators corresponding
to the ground states. From (\ref{eq:Virasoro}), we can see that these
operators are primary fields with conformal weight 
\[
\frac{1}{2}\lambda(1-\lambda)\,.
\]

A ground state can be expressed as a linear combination of 
\[
\left|y\right\rangle
\]
that satisfies 
\begin{eqnarray}
\alpha_{k+\lambda}\left|y\right\rangle  & = & 0\ \ \ \ \ \ \ (k\geq0)\,,\nonumber \\
\tilde{\alpha}_{k+1-\lambda}\left|y\right\rangle  & = & 0\ \ \ \ \ \ \ (k\geq0)\,,\nonumber \\
\frac{1}{\sqrt{2}i}(x-\tilde{x})\left|y\right\rangle  & = & y\left|y\right\rangle \,,\nonumber \\
\left\langle y|y^{\prime}\right\rangle  & = & \delta(y-y^{\prime})\,.\label{eq:yket}
\end{eqnarray}
We can define the BCC operators $\sigma_{*}^{y},\,\bar{\sigma}_{*}^{y}$
such that 
\begin{eqnarray*}
\left|y\right\rangle  & = & \sigma_{*}^{y}(0)\left|0\right\rangle \,,\\
\left\langle y\right| & = & \lim_{z\to\infty}\left\langle 0\right|z^{\lambda(1-\lambda)}\bar{\sigma}_{*}^{y}(z)\,.
\end{eqnarray*}

Another choice of basis of the ground states is given by the eigenstates
of the operator $F_{12}x\tilde{x}$ which is included in the angular
momentum $J$ associated with the rotational symmetry $X\rightarrow e^{i\phi}X,\ \tilde{X}\rightarrow e^{-i\phi}\tilde{X}$:
\begin{equation}
J=F_{12}\,\tilde{x}x-\frac{1}{2}-\sum_{k=0}^{\infty}\left(\frac{1}{k+\lambda}\tilde{\alpha}_{-k-\lambda}\alpha_{k+\lambda}-\frac{1}{k-\lambda+1}\alpha_{-k+\lambda-1}\tilde{\alpha}_{k-\lambda+1}\right),
\end{equation}
where a normal ordering constant is determined by the parity along
the $x^{2}$ direction\footnote{Under the $x^{2}$ parity, the magnetic field $F_{12}$ is transformed
to $-F_{12}$ ($\lambda\rightarrow1-\lambda)$ and the modes are transformed
as 
\[
x\leftrightarrow\tilde{x},\ \ \ \alpha_{k+\lambda}\leftrightarrow\tilde{\alpha}_{k-\lambda+1}.
\]
Since the angular momentum is transformed as $J\rightarrow-J$, the
normal ordering constant turns out to be $1/2$.}. Since the zero modes satisfy $x^{\dagger}=\tilde{x}$ and (\ref{eq:zero_comrel}),
the operator $F_{12}\,\tilde{x}x$ is a kind of number counting operator
and its expectation value must be positive or negative definite,
depending on the signature of $F_{12}$. So, the operator $x$ is
regarded as the annihilation operator and $\tilde{x}$ as the creation
operator if $-F_{12}>0$, and the roles of $x$ and $\tilde{x}$ are
inverted if $-F_{12}<0$. Hence, the vacuum is defined as $x\left|\Omega\right>=0$
for $-F_{12}>0$ or $\tilde{x}\left|\Omega\right>=0$ for $-F_{12}<0$.
We can define the BCC operators $\sigma_{*}^{\Omega},\,\bar{\sigma}_{*}^{\Omega}$
such that 
\begin{eqnarray*}
\left|\Omega\right\rangle  & = & \sigma_{*}^{\Omega}(0)\left|0\right\rangle \,,\\
\left\langle \Omega\right| & = & \lim_{z\to\infty}\left\langle 0\right|z^{\lambda(1-\lambda)}\bar{\sigma}_{*}^{\Omega}(z)\,.
\end{eqnarray*}
The eigenstate of the angular momentum is given as $\tilde{x}^{n}\left|\Omega\right>$
$(n=0,1,2,\cdots)$ for $-F_{12}>0$ and its eigenvalues are $J=-(n+1/2)$.
For $-F_{12}<0$, $x^{n}\left|\Omega\right>$ is the eigenstate of
$J$ corresponding to the eigenstate $n+1/2$. Classically, this is
interpreted as circular motion in a rotational direction for a
fixed direction of the magnetic field and, quantum mechanically, its
angular momentum is discretized.

\subsection{Toroidally compactified theory\label{subsec:Toroidally-compactified-theory}}

We can follow the same procedure as above and deal with the case where
the directions $X^{1},X^{2}$ are toroidally compactified as 
\begin{equation}
X^{1}\sim X^{1}+2\pi R_{1},\ \ \ X^{2}\sim X^{2}+2\pi R_{2}.\label{eq:periodicity}
\end{equation}
We obtain the same mode expansions (\ref{eq:mode_a}), (\ref{eq:mode})
and the commutation relations (\ref{eq:comrel}), (\ref{eq:zero_comrel}).
In this case, we need to introduce two unitary operators 
\begin{equation}
U=\exp\left(i\frac{x^{1}}{R_{1}}\right),\ \ \ V=\exp\left(i\frac{x^{2}}{R_{2}}\right)
\end{equation}
with 
\[
x^{1}=\frac{1}{\sqrt{2}}(x+\tilde{x}),\ \ \ x^{2}=\frac{1}{\sqrt{2}i}(x-\tilde{x})
\]
to get the representation of the zero-mode algebra (\ref{eq:zero_comrel})
consistent with the periodicity (\ref{eq:periodicity}). Since the
Dirac quantization implies 
\begin{equation}
\left(2\pi\right)^{2}R_{1}R_{2}F_{12}=2\pi N\,,\label{eq:DiracQuantization}
\end{equation}
for some integer $N$, we can see that the $U,\,V$ satisfy the relation
\begin{equation}
UV=e^{i\frac{2\pi}{N}}\,VU,\ \ \ (N=\pm1,\pm2,\cdots)\,.\label{eq:UValgebra}
\end{equation}
It is well known that the algebra (\ref{eq:UValgebra}) has 
a $|N|$-dimensional representation. For $\left|N\right|=1$, we can take $U=V=1$.
For $\left|N\right|\ne1$, if we diagonalize the operator $V$, the
representation is explicitly given in matrix form as 
\begin{equation}
U=\begin{bmatrix}0 & 1 & \cdots &  & 0\\
\vdots & 0 & 1 &  & \vdots\\
\vdots &  & \ddots & \ddots & 0\\
0 &  &  & 0 & 1\\
1 & 0 & \cdots & \cdots & 0
\end{bmatrix},\ \ \ V=\begin{bmatrix}1\\
 & \omega\\
 &  & \omega^{2}\\
 &  &  & \ddots\\
 &  &  &  & \omega^{|N|-1}
\end{bmatrix},\ \ \ \ (\omega=e^{i\frac{2\pi}{N}}).
\end{equation}
Let $\left|k\right>$ be the eigenstate of $V$ corresponding to the
eigenvalue $\omega^{k}\ (k\in\mathbb{Z},\,|k|\leq\frac{\left|N\right|}{2})$.
We normalize the eigenstates as 
\begin{equation}
\left<k|l\right>=\delta_{k,\,l}\,.
\end{equation}
We can define the BCC operators $\sigma_{*}^{k},\,\bar{\sigma}_{*}^{k}$
so that 
\begin{eqnarray*}
\left|k\right\rangle  & = & \sigma_{*}^{k}(0)\left|0\right\rangle \,,\\
\left\langle k\right| & = & \lim_{z\to\infty}\left\langle 0\right|z^{\lambda(1-\lambda)}\bar{\sigma}_{*}^{k}(z)\,.
\end{eqnarray*}

The theory discussed in the previous subsection corresponds to the
limit $R_{1},R_{2},|N|\to\infty$ with 
\[
F_{12}=\frac{N}{2\pi R_{1}R_{2}}
\]
fixed. The states $\left|y\right\rangle $ defined in (\ref{eq:yket})
can be given by the limit 
\begin{equation}
\left|y\right\rangle =\lim\left(\sqrt{\frac{|N|}{2\pi R_{2}}}\left|k\right\rangle \right)\,,\label{eq:limit}
\end{equation}
with 
\[
y=\frac{2\pi R_{2}}{N}k
\]
fixed.

\subsection{Correlation functions\label{subsec:Correlation-functions}}

For the construction of Erler-Maccaferri solutions, it is necessary
to obtain the OPEs of the BCC operators. We calculate the three-point
and four-point correlation functions of the BCC operators and derive
the OPE from these.

Let us first consider the theory in the compactified space (\ref{eq:periodicity})
and study the three-point function of the form 
\[
\left<\bar{\sigma}_{*}^{l}(\infty)\,e^{ip\cdot X}(z,\bar{z})\,\sigma_{*}^{k}(0)\right>\,,
\]
which can be expressed as 
\[
\left\langle l\right|e^{ip\cdot X}(z,\bar{z})\left|k\right\rangle
\]
in the operator language. As is demonstrated in appendix \ref{sec:Green's-functions-and},
this correlation function can be evaluated using (\ref{eq:mode}),
(\ref{eq:comrel}), and (\ref{eq:zero_comrel}), and we obtain 
\begin{equation}
\left<\bar{\sigma}_{*}^{l}(\infty)\,e^{ip\cdot X}(z,\bar{z})\,\sigma_{*}^{k}(0)\right>=\begin{cases}
{\displaystyle \frac{\delta^{-\frac{\alpha'p^{2}}{2}}}{|z|^{\alpha'p^{2}}}}\left\langle l\right|e^{i(p\tilde{x}+\tilde{p}x)}\left|k\right\rangle  & (z>0)\\
\\
{\displaystyle \frac{\delta^{-\frac{\alpha'p^{2}}{2}\cos^{2}\pi\lambda}}{|z|^{\alpha'p^{2}\cos^{2}\pi\lambda}}}\left\langle l\right|e^{i(p\tilde{x}+\tilde{p}x)}\left|k\right\rangle  & (z<0)\,.
\end{cases}\label{eq:3ptfunc-torus}
\end{equation}
For 
\[
p_{i}=\frac{n_{i}}{R_{i}}\ \ \ \ \ (i=1,2,\ n_{i}\in\mathbb{Z})\,,
\]
we get 
\[
\left\langle l\right|e^{i(p\tilde{x}+\tilde{p}x)}\left|k\right\rangle =e^{-i\pi\frac{n_{1}n_{2}}{N}}\left\langle l\right|U^{n_{1}}V^{n_{2}}\left|k\right\rangle =\omega^{\frac{n_{1}n_{2}}{2}+n_{2}l}\delta_{k-l,\,n_{1}\,({\rm mod}N)}\,.
\]

With the three-point function (\ref{eq:3ptfunc-torus}), the OPEs
of the operators $\sigma_{*}^{k},\bar{\sigma}_{*}^{l}$ are obtained
as 
\begin{eqnarray}
\sigma_{*}^{k}(s)\bar{\sigma}_{*}^{l}(0) & \sim & \frac{1}{g_{0}}\sum_{n_{1},n_{2}}\left<\bar{\sigma}_{*}^{l}(\infty)\,e^{ip\cdot X}(1,1)\,\sigma_{*}^{k}(0)\right>s^{-\lambda(1-\lambda)+\alpha^{\prime}p^{2}}e^{-ip\cdot X}(0)\,,\nonumber \\
\bar{\sigma}_{*}^{l}(s)\sigma_{*}^{k}(0) & \sim & \frac{1}{g_{*}}\sum_{n_{1},n_{2}}\left<\bar{\sigma}_{*}^{l}(\infty)\,e^{ip\cdot X}(-1,-1)\,\sigma_{*}^{k}(0)\right>s^{-\lambda(1-\lambda)+\alpha^{\prime}p^{2}\cos^{2}\pi\lambda}e^{-ip\cdot X}(0)\,,\label{eq:OPE*}
\end{eqnarray}
for $s>0$. $g_{0}$ is equal to the volume $(2\pi)^{2}R_{1}R_{2}$
of the two-dimensional space, and all we have to do is to obtain $g_{*}$,
which can be derived from the four-point function. The four-point
functions of the BCC operators are calculated in appendix \ref{sec:Green's-function-with},
and we have 
\begin{eqnarray}
 &  & \left<\bar{\sigma}_{*}^{j}(0)\sigma_{*}^{i}(x)\bar{\sigma}_{*}^{k}(1)\sigma_{*}^{l}(\infty)\right>\nonumber \\
 &  & \quad=\frac{1}{\left(2\pi\right)^{2}R_{1}R_{2}}x^{-\lambda(1-\lambda)}(1-x)^{-\lambda(1-\lambda)}\frac{1}{F(\lambda,1-\lambda,1;x)}\,\delta_{j-i,l-k\,{\rm (mod}N)}\nonumber \\
 &  & \hphantom{\quad=}\times\sum_{n,m}\omega^{(j-l)m}\exp\left[-\frac{\pi\alpha'}{\sin\pi\lambda}\frac{F(\lambda,1-\lambda,1;1-x)}{F(\lambda,1-\lambda,1;x)}\left\{ \frac{(j-i+nN)^{2}}{R_{1}^{2}}+\frac{m^{2}}{R_{2}^{2}}\right\} \right]\,.\label{eq:4ptfunc}
\end{eqnarray}
In order to evaluate $g_{*}$, we examine the limit $x\rightarrow1-0$,
which can be studied by rewriting it into the form 
\begin{eqnarray}
 &  & \left<\sigma_{*}^{j}(0)\bar{\sigma}_{*}^{i}(x)\sigma_{*}^{k}(1)\bar{\sigma}_{*}^{l}(\infty)\right>\nonumber \\
 &  & \quad=\frac{|\cos\pi\lambda|}{\left(2\pi\right)^{2}R_{1}R_{2}}\,x^{-\lambda(1-\lambda)}(1-x)^{-\lambda(1-\lambda)}\frac{1}{F(\lambda,1-\lambda,1;1-x)}\,\delta_{j-i,l-k\,{\rm (mod}N)}\nonumber \\
 &  & \hphantom{\quad=}\times\sum_{n,m}\omega^{(i-j)n}\exp\left[-\frac{\pi\alpha'}{\sin\pi\lambda}\cos^{2}\pi\lambda\frac{F(\lambda,1-\lambda,1;x)}{F(\lambda,1-\lambda,1;1-x)}\left\{ \frac{(i-k+mN)^{2}}{R_{1}^{2}}+\frac{n^{2}}{R_{2}^{2}}\right\} \right]\,,\label{eq:4ptposson}
\end{eqnarray}
using the Poisson resummation formula
\begin{align*}
\sum_{m=-\infty}^{\infty}e^{i\frac{2\pi}{N}(j-l)m}e^{-am^{2}} & =\sqrt{\frac{\pi}{a}}\sum_{m=-\infty}^{\infty}e^{-\frac{\pi^{2}}{N^{2}a}(j-l+mN)^{2}}\,,\\
\sum_{n=-\infty}^{\infty}e^{-a(j-i+nN)^{2}} & =\frac{1}{|N|}\sqrt{\frac{\pi}{a}}\sum_{n=-\infty}^{\infty}e^{i\frac{2\pi}{N}(i-j)n}e^{-\frac{\pi^{2}}{N^{2}a}n^{2}}\,.
\end{align*}
From (\ref{eq:4ptposson}), we find that, for $x\sim1$, 
\begin{eqnarray*}
 &  & \left<\bar{\sigma}_{*}^{j}(0)\sigma_{*}^{i}(x)\bar{\sigma}_{*}^{k}(1)\sigma_{*}^{l}(\infty)\right>\\
 &  & \quad\sim\frac{|\cos\pi\lambda|}{\left(2\pi\right)^{2}R_{1}R_{2}}\sum_{\vec{p}}\omega^{n_{1}n_{2}+n_{2}(i-l)}\delta_{k-i,n_{1}\,({\rm mod}N)}\delta_{l-j,n_{1}\,({\rm mod}N)}\delta^{-\alpha'p^{2}}(1-x)^{-\lambda(1-\lambda)+\alpha'p^{2}\cos^{2}\pi\lambda}\,.
\end{eqnarray*}
Comparing this with (\ref{eq:OPE*}), we get 
\begin{equation}
g_{*}=\frac{\left(2\pi\right)^{2}R_{1}R_{2}}{|\cos\pi\lambda|}\,,\label{eq:g*}
\end{equation}
and the OPEs of 
\begin{eqnarray}
 &  & \sigma_{*}^{k}(s)\bar{\sigma}_{*}^{l}(0)\nonumber \\
 &  & \quad\sim\frac{1}{\left(2\pi\right)^{2}R_{1}R_{2}}s^{-\lambda(1-\lambda)}\sum_{n_{1},n_{2}}\left(s\delta^{-\frac{1}{2}}\right)^{\alpha^{\prime}\left(\left(\frac{n_{1}}{R_{1}}\right)^{2}+\left(\frac{n_{2}}{R_{2}}\right)^{2}\right)}\nonumber \\
 &  & \hphantom{\quad\sim\frac{1}{\left(2\pi\right)^{2}R_{1}R_{2}}s^{-\lambda(1-\lambda)}\sum_{n_{1},n_{2}}\quad}\times\omega^{\frac{n_{1}n_{2}}{2}+n_{2}l}\delta_{k-l,\,n_{1}\,({\rm mod}N)}e^{-i\left(\frac{n_{1}}{R_{1}}X^{1}+\frac{n_{2}}{R_{2}}X^{2}\right)}(0)\,,\nonumber \\
 &  & \bar{\sigma}_{*}^{l}(s)\sigma_{*}^{k}(0)\nonumber \\
 &  & \quad\sim\frac{|\cos\pi\lambda|}{\left(2\pi\right)^{2}R_{1}R_{2}}s^{-\lambda(1-\lambda)}\sum_{n_{1},n_{2}}\left(s\delta^{-\frac{1}{2}}\right)^{\alpha^{\prime}\left(\left(\frac{n_{1}}{R_{1}}\right)^{2}+\left(\frac{n_{2}}{R_{2}}\right)^{2}\right)\cos^{2}\pi\lambda}\nonumber \\
 &  & \hphantom{\quad\sim\frac{|\cos\pi\lambda|}{\left(2\pi\right)^{2}R_{1}R_{2}}s^{-\lambda(1-\lambda)}\sum_{n_{1},n_{2}}\quad}\times\omega^{\frac{n_{1}n_{2}}{2}+n_{2}l}\delta_{k-l,\,n_{1}\,({\rm mod}N)}e^{-i\left(\frac{n_{1}}{R_{1}}X^{1}+\frac{n_{2}}{R_{2}}X^{2}\right)}(0)\,.\label{eq:OPElk}
\end{eqnarray}
Since 
\begin{equation}
\frac{1}{|\cos\pi\lambda|}=\sqrt{1+\tan^{2}\pi\lambda}=\sqrt{1+\left(2\pi\alpha^{\prime}F_{12}\right)^{2}}\,,\label{eq:cospilambda}
\end{equation}
$g_{*}$ can be expressed as 
\[
g_{*}=\left(2\pi\right)^{2}R_{1}R_{2}\sqrt{\det\left(\begin{array}{cc}
1 & 2\pi\alpha^{\prime}F_{12}\\
2\pi\alpha^{\prime}F_{21} & 1
\end{array}\right)}=\int d^{2}x\sqrt{\det\left(\begin{array}{cc}
1 & 2\pi\alpha^{\prime}F_{12}\\
2\pi\alpha^{\prime}F_{21} & 1
\end{array}\right)}\,,
\]
which coincides with the contribution to the Born-Infeld action from
the two dimensions we are dealing with.

The noncompact case can be dealt with in the same way. The three-point
function and the four-point function of $\sigma^{\Omega},\bar{\sigma}^{\Omega}$
can be calculated to be 
\begin{equation}
\left<\sigma_{*}^{\Omega}(\infty)\,e^{ip\cdot X}(z,\bar{z})\,\bar{\sigma}_{*}^{\Omega}(0)\right>=\begin{cases}
{\displaystyle \frac{\delta^{-\frac{\alpha'p^{2}}{2}}}{|z|^{\alpha'p^{2}}}}\exp\left(-\frac{p^{2}}{4|F_{12}|}\right) & (z>0)\\
\\
{\displaystyle \frac{\delta^{-\frac{\alpha'p^{2}}{2}\cos^{2}\pi\lambda}}{|z|^{\alpha'p^{2}\cos^{2}\pi\lambda}}}\exp\left(-\frac{p^{2}}{4|F_{12}|}\right) & (z<0)\,,
\end{cases}\label{eq:3ptfunc-1}
\end{equation}
\begin{eqnarray}
 &  & \left<\bar{\sigma}_{*}^{\Omega}(0)\sigma_{*}^{\Omega}(x)\bar{\sigma}_{*}^{\Omega}(1)\sigma_{*}^{\Omega}(\infty)\right>\nonumber \\
 &  & \quad=x^{-\lambda(1-\lambda)}(1-x)^{-\lambda(1-\lambda)}\frac{1}{F(\lambda,1-\lambda,1;x)}\nonumber \\
 &  & \hphantom{\quad=}\times\int\frac{d^{2}p}{(2\pi)^{2}}\exp\left[-\frac{\pi\alpha'}{|\tan\pi\lambda|}p^{2}-\frac{\pi\alpha'}{\sin\pi\lambda}\frac{F(\lambda,1-\lambda,1;1-x)}{F(\lambda,1-\lambda,1;x)}p^{2}\right]\nonumber \\
 &  & \quad=|\cos\pi\lambda|\,x^{-\lambda(1-\lambda)}(1-x)^{-\lambda(1-\lambda)}\frac{1}{F(\lambda,1-\lambda,1;1-x)}\nonumber \\
 &  & \hphantom{\quad=}\times\int\frac{d^{2}p}{(2\pi)^{2}}\exp\left[-\frac{\pi\alpha'}{|\tan\pi\lambda|}p^{2}-\frac{\pi\alpha'}{\sin\pi\lambda}\cos^{2}\pi\lambda\frac{F(\lambda,1-\lambda,1;x)}{F(\lambda,1-\lambda,1;1-x)}p^{2}\right].\label{eq:4ptfunc-infty}
\end{eqnarray}
For $z\in\mathbb{R}$, the two-point functions of the $\mbox{BCFT}_{0}$
on the upper half plane are normalized as 
\begin{equation}
\left\langle e^{ip\cdot X}(z,\bar{z})e^{ip^{\prime}\cdot X}(0,0)\right\rangle _{\mbox{BCFT}_{0}}=(2\pi)^{2}\delta^{2}(p+p^{\prime})\left|z\right|^{-2\alpha'p^{2}}\,,\label{eq:BCFT0}
\end{equation}
and those for the $\mbox{BCFT}_{*}$ should be 
\begin{equation}
\left\langle e^{ip\cdot X}(z,\bar{z})e^{ip^{\prime}\cdot X}(0,0)\right\rangle _{\mbox{BCFT}_{*}}=\frac{1}{|\cos\pi\lambda|}(2\pi)^{2}\delta^{2}(p+p^{\prime})\left|z\right|^{-2\alpha'p^{2}\cos^{2}\pi\lambda}\,.\label{eq:BCFTstar}
\end{equation}
Equations (\ref{eq:BCFT0}) and (\ref{eq:BCFTstar}) imply 
\begin{eqnarray*}
g_{0} & = & \left\langle 1\right\rangle _{\mbox{BCFT}_{0}}=\left.(2\pi)^{2}\delta^{2}(p+p^{\prime})\right|_{p=p^{\prime}=0}=\int d^{2}x\,,\\
g_{*} & = & \left\langle 1\right\rangle _{\mbox{BCFT}_{*}}=\int d^{2}x\sqrt{\det\left(\begin{array}{cc}
1 & 2\pi\alpha^{\prime}F_{12}\\
2\pi\alpha^{\prime}F_{21} & 1
\end{array}\right)}\,,
\end{eqnarray*}
although both $g_{0}$ and $g_{*}$ are infinite. We can derive the
OPE 
\begin{eqnarray}
\sigma_{*}^{\Omega}(s)\bar{\sigma}_{*}^{\Omega}(0) & \sim & \int\frac{d^{2}p}{(2\pi)^{2}}\left<\bar{\sigma}_{*}^{\Omega}(\infty)\,e^{ip\cdot X}(1,1)\,\sigma_{*}^{\Omega}(0)\right>s^{-\lambda(1-\lambda)+\alpha^{\prime}p^{2}}e^{-ip\cdot X}(0)\nonumber \\
 & = & s^{-\lambda(1-\lambda)}\int\frac{d^{2}p}{(2\pi)^{2}}\left(s\delta^{-\frac{1}{2}}\right)^{\alpha'p^{2}}e^{-\frac{\pi\alpha'}{2|\tan\pi\lambda|}p^{2}}e^{ip\cdot X}(0)\,,\nonumber \\
\bar{\sigma}_{*}^{\Omega}(s)\sigma_{*}^{\Omega}(0) & \sim & |\cos\pi\lambda|\int\frac{d^{2}p}{(2\pi)^{2}}\left<\bar{\sigma}_{*}^{\Omega}(\infty)\,e^{ip\cdot X}(-1,-1)\,\sigma_{*}^{\Omega}(0)\right>s^{-\lambda(1-\lambda)+\alpha^{\prime}p^{2}\cos^{2}\pi\lambda}e^{-ip\cdot X}(0)\nonumber \\
 & \sim & |\cos\pi\lambda|\,s^{-\lambda(1-\lambda)}\int\frac{d^{2}p}{(2\pi)^{2}}\left(s\delta^{-\frac{1}{2}}\right)^{\alpha'p^{2}\cos^{2}\pi\lambda}e^{-\frac{\pi\alpha'}{2|\tan\pi\lambda|}p^{2}}e^{ip\cdot X}(0)
\label{eq:OPEsigmaomega}
\end{eqnarray}
for $s>0$. They are given as an integral of the exponential operator
over the continuous momenta and it is difficult to construct BCC operators
satisfying (\ref{eq:OPEsigma}) from these.

OPEs of $\sigma_{*}^{y},\bar{\sigma}_{*}^{y}$ can be obtained either
by calculating the correlation functions or simply by taking the limit
(\ref{eq:limit}) of (\ref{eq:OPElk}). We get 
\begin{eqnarray}
 &  & \sigma_{*}^{y}(s)\bar{\sigma}_{*}^{y^{\prime}}(0)\nonumber \\
 &  & \quad\sim s^{-\lambda(1-\lambda)}\frac{\left|\tan\pi\lambda\right|}{2\pi\alpha^{\prime}}\int\frac{dp_{2}}{(2\pi)^{2}}\left(s\delta^{-\frac{1}{2}}\right)^{\alpha^{\prime}\left(\left(\frac{\tan\pi\lambda}{2\pi\alpha^{\prime}}(y-y^{\prime})\right)^{2}+\left(p_{2}\right)^{2}\right)}\nonumber \\
 &  & \hphantom{\quad\sim s^{-\lambda(1-\lambda)}\frac{\left|\tan\pi\lambda\right|}{2\pi\alpha^{\prime}}\int\frac{dp_{2}}{(2\pi)^{2}}\quad}\times e^{ip_{2}\frac{y+y^{\prime}}{2}}e^{-i\left(\frac{\tan\pi\lambda}{2\pi\alpha^{\prime}}(y-y^{\prime})X^{1}+p_{2}X^{2}\right)}(0)\,,\nonumber \\
 &  & \bar{\sigma}_{*}^{y^{\prime}}(s)\sigma_{*}^{y}(0)\nonumber \\
 &  & \quad\sim s^{-\lambda(1-\lambda)}\frac{\left|\sin\pi\lambda\right|}{2\pi\alpha^{\prime}}\int\frac{dp_{2}}{(2\pi)^{2}}\left(s\delta^{-\frac{1}{2}}\right)^{\alpha^{\prime}\left(\left(\frac{\tan\pi\lambda}{2\pi\alpha^{\prime}}(y-y^{\prime})\right)^{2}+\left(p_{2}\right)^{2}\right)\cos^{2}\pi\lambda}\nonumber \\
 &  & \hphantom{\quad\sim s^{-\lambda(1-\lambda)}\frac{\left|\sin\pi\lambda\right|}{2\pi\alpha^{\prime}}\int\frac{dp_{2}}{(2\pi)^{2}}\quad}\times e^{ip_{2}\frac{y+y^{\prime}}{2}}e^{-i\left(\frac{\tan\pi\lambda}{2\pi\alpha^{\prime}}(y-y^{\prime})X^{1}+p_{2}X^{2}\right)}(0)\label{eq:OPEsigmay}
\end{eqnarray}
for $s>0$.

\section{Classical solutions for constant magnetic field\label{sec:Classical-solutions-for}}

With the OPEs (\ref{eq:OPElk}), (\ref{eq:OPEsigmaomega}), and (\ref{eq:OPEsigmay})
derived in the previous section, we are able to construct the Erler-Maccaferri
solution corresponding to the constant magnetic field background.
All we have to do is to find BCC operators $\sigma_{*},\,\bar{\sigma}_{*}$
satisfying (\ref{eq:sigmastar}).

\subsection{Toroidally compactified theory}

In the case of the theory in the toroidally compactified space, it
is straightforward to construct $\sigma_{*},\,\bar{\sigma}_{*}$.
Since (\ref{eq:OPElk}) imply 
\begin{eqnarray*}
\sigma_{*}^{k}(s)\bar{\sigma}_{*}^{l}(0) & \sim & \frac{1}{\left(2\pi\right)^{2}R_{1}R_{2}}s^{-\lambda(1-\lambda)}\delta_{k,\,l}\,,\\
\bar{\sigma}_{*}^{l}(s)\sigma_{*}^{k}(0) & \sim & \frac{|\cos\pi\lambda|}{\left(2\pi\right)^{2}R_{1}R_{2}}s^{-\lambda(1-\lambda)}\delta_{k,\,l}
\end{eqnarray*}
for small positive $s$, one can take 
\begin{eqnarray}
\sigma_{*}(s) & = & \sqrt{\frac{\left(2\pi\right)^{2}R_{1}R_{2}}{|\cos\pi\lambda|}}\sigma_{*}^{k}(s)\,,\nonumber \\
\bar{\sigma}_{*}(s) & = & \sqrt{\frac{\left(2\pi\right)^{2}R_{1}R_{2}}{|\cos\pi\lambda|}}\bar{\sigma}_{*}^{k}(s)
\label{eq:F12sigmastar}
\end{eqnarray}
for some $k\,(\left|k\right|\leq\frac{|N|}{2})$ which satisfy the
OPE (\ref{eq:sigmastar}), with 
\[
h=\frac{1}{2}\lambda(1-\lambda),\ \ \ \ \ \frac{g_{*}}{g_{0}}=\frac{1}{|\cos\pi\lambda|}\,.
\]
From these, one can construct the Erler-Maccaferri solution (\ref{eq:EelerMaccaferri})
which describes the D-branes with gauge field strength $F_{12}$.
From (\ref{eq:cospilambda}), we can see that the action of the background
is given by the difference between the Born-Infeld action and the
D-brane tension.

It is also possible to construct solutions corresponding to multiple
brane solutions with the $F_{12}$ background by considering 
\begin{eqnarray}
\sigma_{*,k}(s) & = & \sqrt{\frac{\left(2\pi\right)^{2}R_{1}R_{2}}{|\cos\pi\lambda|}}\sigma_{*}^{k}(s)\,,\nonumber \\
\bar{\sigma}_{*,l}(s) & = & \sqrt{\frac{\left(2\pi\right)^{2}R_{1}R_{2}}{|\cos\pi\lambda|}}\bar{\sigma}_{*}^{l}(s)\,,\label{eq:F12sigmastark}
\end{eqnarray}
from which we can construct string fields $\Sigma_{k},\bar{\Sigma}_{l}$
similarly to (\ref{eq:Sigma}) satisfying
\begin{equation}
\bar{\Sigma}_{k}\Sigma_{l}=\delta_{k,\,l}\,,
\end{equation}
so that 
\begin{equation}
\Psi=\Psi_{{\rm tv}}-\sum_{k=1}^{M}\Sigma_{k}\Psi_{{\rm tv}}\bar{\Sigma}_{k}
\end{equation}
gives a solution for $0<M\leq|N|$. This solution can be regarded
as a solution corresponding to $M$ D-branes with magnetic field condensation.\footnote{ More general solutions can be given by 
\[
\Psi=\Psi_{{\rm tv}}-\sum_{i,j}A_{ij}\Sigma_{i}\Psi_{{\rm tv}}\bar{\Sigma}_{j}\,,
\]
where $A_{ij}$ denotes a Hermitian $|N|\times|N|$ matrix satisfying
\[
A^{2}=A\,.
\]
$A$ can be considered as a projection operator. } $M$ cannot be greater than $|N|$ because the BCC operators we consider
in this paper are those corresponding to the ground states of the
Fock space in subsection \ref{subsec:First-quantization section}.
It should be possible to deal with more general cases by introducing
BCC operators corresponding to other states.

It is also possible to construct solutions with general $F_{\mu\nu}$
of the form (\ref{eq:block}) by considering the tensor product of
the BCC operators (\ref{eq:F12sigmastark}), one for each block. It
is straightforward to show that the energy of the solution is given
by the difference of the Born-Infeld action and the D-brane tension.

\subsection{Noncompact case}

Unlike the compactified case, (\ref{eq:OPEsigmay}) and (\ref{eq:OPEsigmaomega})
include extra $\ln s$ dependence for small positive $s$, and therefore
we cannot choose $\sigma_{*}^{y},\,\bar{\sigma}_{*}^{y}$ or $\sigma_{*}^{\Omega},\,\bar{\sigma}_{*}^{\Omega}$
as the $\sigma_{*},\,\bar{\sigma}_{*}$ satisfying the OPE (\ref{eq:sigmastar}).
One somewhat artificial way to construct such BCC operators is to
make the following linear combinations of $\sigma_{*}^{y},\,\bar{\sigma}_{*}^{y}$:
\begin{eqnarray*}
\sigma_{*}(s) & \equiv & \sqrt{\frac{(2\pi)^{2}a\alpha^{\prime}}{\left|\sin\pi\lambda\right|}}\sum_{n=-\infty}^{\infty}\sigma_{*}^{na}(s)\,,\\
\bar{\sigma}_{*}(s) & \equiv & \sqrt{\frac{(2\pi)^{2}a\alpha^{\prime}}{\left|\sin\pi\lambda\right|}}\sum_{n=-\infty}^{\infty}\sigma_{*}^{na}(s)
\end{eqnarray*}
for $a>0$. It is straightforward to derive the following OPEs for
$s>0$ from (\ref{eq:OPEsigmay}): 
\begin{eqnarray*}
\bar{\sigma}_{*}(s)\sigma_{*}(0) & \sim & s^{-\lambda(1-\lambda)}\,,\\
\sigma_{*}(s)\bar{\sigma}_{*}(0) & \sim & \frac{1}{\left|\cos\pi\lambda\right|}s^{-\lambda(1-\lambda)}\,.
\end{eqnarray*}
From $\sigma_{*},\,\bar{\sigma}_{*}$, we can construct solutions
corresponding to the D-brane with the $F_{\mu\nu}$ background in the
way explained in the previous subsection. Similarly, by taking different
linear combinations of $\sigma_{*}^{y},\,\bar{\sigma}_{*}^{y}$, such
as 
\begin{eqnarray*}
\sigma_{*,k}(s) & \equiv & \sqrt{\frac{(2\pi)^{2}Ma\alpha^{\prime}}{\left|\sin\pi\lambda\right|}}\sum_{n=-\infty}^{\infty}\sigma_{*}^{(nM+k)a}(s)\,,\\
\bar{\sigma}_{*,l}(s) & \equiv & \sqrt{\frac{(2\pi)^{2}Ma\alpha^{\prime}}{\left|\sin\pi\lambda\right|}}\sum_{n=-\infty}^{\infty}\sigma_{*}^{(nM+l)a}(s)
\end{eqnarray*}
($M\in\mathbb{\mathbb{N}};\,k,l=0,1,\cdots,M-1$), we have 
\begin{eqnarray*}
\bar{\sigma}_{*,k}(s)\sigma_{*,l}(0) & \sim & s^{-\lambda(1-\lambda)}\delta_{k,l}\,,\\
\sigma_{*,k}(s)\bar{\sigma}_{*,l}(0) & \sim & \frac{1}{\left|\cos\pi\lambda\right|}s^{-\lambda(1-\lambda)}\delta_{k,l}
\end{eqnarray*}
for $s>0$ and we can construct solutions corresponding to multiple
D-branes with $F_{\mu\nu}$ from them.

\section{Concluding remarks\label{sec:Concluding-remarks}}

In this paper, we have constructed the Erler-Maccaferri solutions
corresponding to constant magnetic field configurations on D-branes.
In order to do so, we have calculated the correlation functions of
the BCC operators and obtained their OPEs of them in the cases of the
theories in toroidally compactified and noncompact space. We have
shown that the Born-Infeld action appears as the action for such backgrounds.

There are several important things to be studied about the solutions
obtained in this paper. In the case of toroidally compactified space,
the Chern number of the $U(1)$ gauge field is nonvanishing. Therefore
the configuration should be a topologically nontrivial one, from the
low energy point of view. One question is how such configurations
are realized as those of the string field. Such a question may be
studied by examining the position space profile of the gauge field,
as was done in \cite{Erler2014}. From the OPE of the operators $\sigma_{*},\,\bar{\sigma}_{*}$,
one can calculate the profile, which is easily seen to be a periodic
function of the coordinates $x^{1},x^{2}$. Therefore one expects
that the gauge field profile has jump discontinuities as a function
of $x^{1},x^{2}$ in order for the configuration to be topologically
nontrivial\footnote{We thank T. Erler, M. Schnabl, and M. Kudrna for pointing this out to the
authors. }. We leave this for future work.

With the solution constructed in this paper, we should be able to
deduce all the interesting features of the open string theory around
the background. In particular, the relation to the noncommutative geometry
should be seen by analyzing the solution. One thing that can be done
would be to get the relation between the worldsheet variables of $\mbox{BCFT}_{0}$
and $\mbox{BCFT}_{*}$ by using the string field theory technique.
In \cite{Erler2014}, it is pointed out that the correspondence between
the states in $\mbox{BCFT}_{0}$ and $\mbox{BCFT}_{*}$ can be given
by the string fields $\Sigma$, $\bar{\Sigma}$. It will be interesting
to explore such a correspondence and compare with the one given in
\cite{Sugino2000, Kawano2000a}.

\section*{Acknowledgments}

We would like to thank T. Erler, M. Kudrna, C. Maccaferri, and M. Schnabl
for useful comments. We also would like to thank the organizers of
``VIII Workshop on String Field Theory and Related Aspects'' at
S\~ao Paulo, especially N. Berkovits, for hospitality. This work was
supported in part by JSPS Grant-in-Aid for Scientific Research (C)
(JP25400242), JSPS Grant-in-Aid for Young Scientists (B) (JP25800134),
and JSPS Grant-in-Aid for Scientific Research (C) (JP15K05056).

\appendix

\section{First quantization of open strings in background gauge field \label{sec:First-quantization-of}}

In this appendix, we discuss the first quantization of open strings
in background gauge fields. It has been discussed in \cite{Abouelsaood1987}
and we present their results here in order to fix our notation.

Let us consider the worldsheet theory of the variables $X,\tilde{X}$
defined in subsection \ref{subsec:First-quantization section}. Here,
we study the general case where the charges at the ends of the strings
are $q_{1},q_{2}$. The action is

\begin{equation}
S=\frac{1}{2\pi\alpha'}\int dtd\sigma(\dot{X}\dot{\tilde{X}}-X'\tilde{X}')-\frac{i\theta}{4\pi\alpha'}\int dt\left[q_{1}(X\dot{\tilde{X}}-\tilde{X}\dot{X})\Big|_{\sigma=0}+q_{2}(X\dot{\tilde{X}}-\tilde{X}\dot{X})\Big|_{\sigma=\pi}\right]\,,
\end{equation}
where $\theta=-2\pi\alpha'F_{12}$. The system discussed in subsection
\ref{subsec:First-quantization section} corresponds to the case $q_{1}=0,\,q_{2}=1$.
The canonical momentum is defined as 
\begin{align}
P(t,\sigma) & =\frac{\partial{\cal L}}{\partial\dot{X}}=\frac{1}{2\pi\alpha'}\dot{\tilde{X}}+\frac{i\theta}{4\pi\alpha'}\left\{ q_{1}\delta(\sigma)+q_{2}\delta(\pi-\sigma)\right\} \tilde{X}\,,\\
\tilde{P}(t,\sigma) & =\frac{\partial{\cal L}}{\partial\dot{\tilde{X}}}=\frac{1}{2\pi\alpha'}\dot{X}-\frac{i\theta}{4\pi\alpha'}\left\{ q_{1}\delta(\sigma)+q_{2}\delta(\pi-\sigma)\right\} X\,,
\end{align}
and they satisfy the canonical commutation relations 
\begin{equation}
\left[X(t,\sigma),\,P(t,\sigma')\right]=\left[\tilde{X}(t,\sigma),\,\tilde{P}(t,\sigma')\right]=i\delta(\sigma-\sigma')\,,
\end{equation}
with all other commutators vanishing. The boundary conditions are
given by 
\begin{align}
X' & =-iq_{1}\theta\dot{X}\ \ \ (\sigma=0),\ \ \ X'=iq_{2}\theta\dot{X}\ \ \ (\sigma=\pi)\,,\label{eq:bcA1}\\
\tilde{X}' & =iq_{1}\theta\dot{\tilde{X}}\ \ \ (\sigma=0),\ \ \ \tilde{X}'=-iq_{2}\theta\dot{\tilde{X}}\ \ \ (\sigma=\pi)\,.\label{eq:bcA2}
\end{align}

First we consider the case $q_{1}+q_{2}\neq0$. The variables $X,\tilde{X}$
can be expanded as 
\begin{align}
X(t,\sigma) & =x+i\sqrt{2\alpha'}\sum_{k=-\infty}^{\infty}\frac{1}{k+\lambda_{1}+\lambda_{2}}\alpha_{k+\lambda_{1}+\lambda_{2}}\psi_{k}(t,\sigma)\,,\\
\tilde{X}(t,\sigma) & =\tilde{x}+i\sqrt{2\alpha'}\sum_{k=-\infty}^{\infty}\frac{1}{k-\lambda_{1}-\lambda_{2}}\tilde{\alpha}_{k-\lambda_{1}-\lambda_{2}}\psi_{-k}^{*}(t,\sigma)\,,
\end{align}
where $q_{i}\theta=-\tan\pi\lambda_{i}\ (i=1,2)$ . $\psi_{k}(t,\sigma)$
are the mode functions which are solutions to the wave equation with
boundary conditions (\ref{eq:bcA1}): 
\begin{equation}
\psi_{k}(t,\sigma)=e^{-i(k+\lambda_{1}+\lambda_{2})t}\cos\left[(k+\lambda_{1}+\lambda_{2})\sigma-\pi\lambda_{1}\right]\,.
\end{equation}
Their complex conjugates $\psi_{k}^{*}(t,\sigma)$ satisfy the boundary
conditions (\ref{eq:bcA2}). Since $\tilde{X}$ is the Hermitian conjugate
of $X$, the operators $x,\tilde{x},\alpha_{k+\lambda_{1}+\lambda_{2}}$, and $\tilde{\alpha}_{k+\lambda_{1}+\lambda_{2}}$
obey 
\begin{equation}
x^{\dagger}=\tilde{x}\,,\ \ \ (\alpha_{k+\lambda_{1}+\lambda_{2}})^{\dagger}=\tilde{\alpha}_{-k-\lambda_{1}-\lambda_{2}}\,.
\end{equation}

The orthogonality relations are given by 
\begin{align}
\int_{0}^{\pi}d\sigma\psi_{k}^{*}(t,\sigma)D\psi_{l}(t,\sigma) & =\pi(k+\lambda_{1}+\lambda_{2})\delta_{k,\,l}\,,\\
\int_{0}^{\pi}d\sigma D\psi_{k}(t,\sigma) & =0\,,
\end{align}
where $D=i\stackrel{\rightarrow}{\partial_{t}}-i\stackrel{\leftarrow}{\partial_{t}}+q_{1}\theta\delta(\sigma)+q_{2}\theta\delta(\pi-\sigma).$
Since $\psi_{k}(t,\sigma)$ and the constant mode form a complete
set, the operators $x,\tilde{x},\alpha_{k+\lambda_{1}+\lambda_{2}}$, and $\tilde{\alpha}_{k+\lambda_{1}+\lambda_{2}}$
are expressed as 
\begin{align}
\alpha_{k+\lambda_{1}+\lambda_{2}} & =\frac{1}{\sqrt{2\alpha'}\pi i}\int_{0}^{\pi}d\sigma\,\psi_{k}^{*}\left[2\pi\alpha'i\tilde{P}+\left\{ k+\lambda_{1}+\lambda_{2}+\frac{q_{1}\theta}{2}\delta(\sigma)+\frac{q_{2}\theta}{2}\delta(\pi-\sigma)\right\} X\right]\,,\\
\tilde{\alpha}_{k-\lambda_{1}-\lambda_{2}} & =\frac{1}{\sqrt{2\alpha'}\pi i}\int_{0}^{\pi}d\sigma\,\psi_{-k}\left[2\pi\alpha'iP+\left\{ k-\lambda_{1}-\lambda_{2}-\frac{q_{1}\theta}{2}\delta(\sigma)-\frac{q_{2}\theta}{2}\delta(\pi-\sigma)\right\} \tilde{X}\right]\,,\\
x & =\frac{1}{(q_{1}+q_{2})\theta}\int_{0}^{\pi}d\sigma\,\left[2\pi\alpha'i\tilde{P}+\left\{ \frac{q_{1}\theta}{2}\delta(\sigma)+\frac{q_{2}\theta}{2}\delta(\pi-\sigma)\right\} X\right]\,,\\
\tilde{x} & =\frac{1}{(q_{1}+q_{2})\theta}\int_{0}^{\pi}d\sigma\,\left[-2\pi\alpha'iP+\left\{ \frac{q_{1}\theta}{2}\delta(\sigma)+\frac{q_{2}\theta}{2}\delta(\pi-\sigma)\right\} \tilde{X}\right]\,.
\end{align}
From these expressions and the canonical commutation relations, it
follows that 
\begin{equation}
\left[\alpha_{k+\lambda_{1}+\lambda_{2}},\,\tilde{\alpha}_{l-\lambda_{1}-\lambda_{2}}\right]=(k+\lambda_{1}+\lambda_{2})\delta_{k+l,0}\,,\ \ \ \left[x,\,\tilde{x}\right]=\frac{2\pi\alpha'}{(q_{1}+q_{2})\theta}\,.
\end{equation}
The energy-momentum tensor is defined as 
\[
T(z)=\lim_{z'\rightarrow z}\left[-\frac{2}{\alpha'}\,\partial X(z)\partial\tilde{X}(z')-\frac{1}{(z-z')^{2}}\right]\,,
\]
and we get the Virasoro generator 
\[
L_{n}=\sum_{k=-\infty}^{\infty}:\alpha_{n-k+\lambda_{1}+\lambda_{2}}\tilde{\alpha}_{k-\lambda_{1}-\lambda_{2}}:+\frac{1}{2}(\lambda_{1}+\lambda_{2})(1-\lambda_{1}-\lambda_{2})\delta_{n,0}\,.
\]

Let us turn to the case $q_{1}+q_{2}=0$, which corresponds to a neutral
string. An easy way to deal with this case is to take the limit $\lambda_{1}=-\lambda+\epsilon,\,\lambda_{2}=\lambda+\epsilon,\,\epsilon\to0$
of the above results. Although the limit of the zero modes requires
some care, we finally obtain the mode expansions 
\begin{align}
X(t,\sigma) & =x_{0}+\sqrt{2\alpha'}\alpha_{0}\left\{ t\cos\pi\lambda-i\left(\sigma-\frac{\pi}{2}\right)\sin\pi\lambda\right\} \nonumber \\
 & \ +i\sqrt{2\alpha'}\sum_{n\neq0}\frac{1}{n}\alpha_{n}e^{-int}\cos(n\sigma+\pi\lambda)\,,\label{eq:X}\\
\tilde{X}(t,\sigma) & =\tilde{x}_{0}+\sqrt{2\alpha'}\tilde{\alpha}_{0}\left\{ t\cos\pi\lambda+i\left(\sigma-\frac{\pi}{2}\right)\sin\pi\lambda\right\} \nonumber \\
 & \ +i\sqrt{2\alpha'}\sum_{n\neq0}\frac{1}{n}\tilde{\alpha}_{n}e^{-int}\cos(n\sigma-\pi\lambda)\label{eq:Xtilde}
\end{align}
and the commutation relations 
\begin{equation}
\left[x_{0},\,\tilde{\alpha}_{0}\right]=\left[\tilde{x}_{0,}\,\alpha_{0}\right]=i\sqrt{2\alpha'}\cos\pi\lambda\,,\ \ \ \left[\alpha_{m},\,\tilde{\alpha}_{m}\right]=n\delta_{n+m,\,0}\,,\label{eq:ccr_neutsting}
\end{equation}
with all other commutators vanishing. The Virasoro generators are
given by 
\[
L_{n}=\sum_{k}:\alpha_{n-k}\tilde{\alpha}_{k}:\,.
\]
The Fock vacuum $\left|\vec{p}\right\rangle $ can be defined to satisfy
\begin{eqnarray*}
\alpha_{0}\left|\vec{p}\right\rangle  & = & \sqrt{2\alpha'}p\cos\pi\lambda\left|\vec{p}\right\rangle \,,\\
\tilde{\alpha}_{0}\left|\vec{p}\right\rangle  & = & \sqrt{2\alpha'}\tilde{p}\cos\pi\lambda\left|\vec{p}\right\rangle \,.
\end{eqnarray*}
$\left|\vec{p}\right\rangle $ corresponds to the primary field $e^{i\vec{p}\cdot\vec{X}}(z,\bar{z})=e^{i(\tilde{p}X+p\tilde{X})}(z,\bar{z})$
with weight $\alpha'p^{2}\cos^{2}\pi\lambda$ 
\[
\left|\vec{p}\right\rangle =e^{i\vec{p}\cdot\vec{X}}(0,0)\left|0\right\rangle \,.
\]
The operator $e^{i\vec{p}\cdot\vec{X}}(0,0)$ here is normal ordered
and can be expressed more precisely as 
\begin{eqnarray}
e^{i\vec{p}\cdot\vec{X}}(0,0) & = & \lim_{\epsilon\to0}\left\{ \exp\left[\frac{i}{\pi}\int_{0}^{\pi}d\theta\left(p\tilde{X}(\epsilon e^{i\theta},\epsilon e^{-i\theta})+\tilde{p}X(\epsilon e^{i\theta},\epsilon e^{-i\theta})\right)\right]\vphantom{\left[-\frac{\alpha^{\prime}p\tilde{p}}{\pi^{2}}\int_{0}^{\pi}d\theta\int_{0}^{\pi}d\theta^{\prime}\left(\ln\left|\epsilon e^{i\theta}-\epsilon e^{i\theta^{\prime}}\right|+\cos2\pi\lambda\ln\left|\epsilon e^{i\theta}-\epsilon e^{-i\theta^{\prime}}\right|\right)\right]}\right.\nonumber \\
 &  & \hphantom{\lim_{\epsilon\to0}\exp}\times\left.\exp\left[-\frac{\alpha^{\prime}p\tilde{p}}{\pi^{2}}\int_{0}^{\pi}d\theta\int_{0}^{\pi}d\theta^{\prime}\left(\ln\left|\epsilon e^{i\theta}-\epsilon e^{i\theta^{\prime}}\right|+(\cos2\pi\lambda)\ln\left|\epsilon e^{i\theta}-\epsilon e^{-i\theta^{\prime}}\right|\right)\right]\right\} \,.\nonumber \\
\label{eq:precise}
\end{eqnarray}
Here, for regularization, we replace the local operator $X(z,\bar{z})$
by an integral along the small contour around $z=0$. Taking the limit
$\epsilon\to0$ with the factor on the second line of (\ref{eq:precise}),
we get the operator $e^{i\vec{p}\cdot\vec{X}}(0,0)$ normal ordered.
This expression is useful in the calculation of three-point functions.

\section{Three-point functions\label{sec:Green's-functions-and}}

In this appendix, we show how to calculate the correlation functions
of the form 
\begin{equation}
\left\langle \ \right|e^{i\vec{p}\cdot\vec{X}}(z,\bar{z})\left|\ \right\rangle ^{\prime}\label{eq:three-point}
\end{equation}
for $z\in\mathbb{R}$, which play crucial roles in deriving the OPE
of the BCC operators in subsection \ref{subsec:Correlation-functions}.
Here, $\left|\ \right\rangle ^{\prime},\,\left\langle \ \right|$ are
states in the Fock space defined in subsections \ref{subsec:First-quantization section}
and \ref{subsec:Toroidally-compactified-theory}, and satisfy 
\begin{eqnarray}
\alpha_{k+\lambda}\left|\ \right\rangle ^{\prime} & = & 0\,,\nonumber \\
\tilde{\alpha}_{k+1-\lambda}\left|\ \right\rangle ^{\prime} & = & 0\,,\nonumber \\
\left\langle \ \right|\alpha_{-k-1+\lambda} & = & 0\,,\nonumber \\
\left\langle \ \right|\tilde{\alpha}_{-k-\lambda} & = & 0\label{eq:ground}
\end{eqnarray}
for $k\geq0$. Such correlation functions can be evaluated by essentially
following the method in \cite{Mukhopadhyay2001}.

For $z>0$, we take the primary field $e^{i\vec{p}\cdot\vec{X}}$
in (\ref{eq:three-point}) to be the one corresponding to the delta
function normalized ground state in the $\mbox{BCFT}_{0}$. Therefore
it should coincide with (\ref{eq:precise}) in the $\lambda=0$ case,
and we get 
\begin{eqnarray}
 &  & \left\langle \ \right|e^{i\vec{p}\cdot\vec{X}}(z,\bar{z})\left|\ \right\rangle ^{\prime}\nonumber \\
 &  & \quad=\lim_{\epsilon\to0}\left\{ \left\langle \ \right|\mathrm{R}\exp\left[\frac{i}{\pi}\int_{0}^{\pi}d\theta\left(p\tilde{X}(z+\epsilon e^{i\theta},\bar{z}+\epsilon e^{-i\theta})+\tilde{p}X(z+\epsilon e^{i\theta},\bar{z}+\epsilon e^{-i\theta})\right)\right]\left|\ \right\rangle ^{\prime}\vphantom{\left[-\frac{\alpha^{\prime}p\tilde{p}}{\pi^{2}}\int_{0}^{\pi}d\theta\int_{0}^{\pi}d\theta^{\prime}\left(\ln\left|\epsilon e^{i\theta}-\epsilon e^{i\theta^{\prime}}\right|+\ln\left|\epsilon e^{i\theta}-\epsilon e^{-i\theta^{\prime}}\right|\right)\right]}\right.\nonumber \\
 &  & \quad\hphantom{=\lim_{\epsilon\to0}\exp}\times\left.\exp\left[-\frac{\alpha^{\prime}p\tilde{p}}{\pi^{2}}\int_{0}^{\pi}d\theta\int_{0}^{\pi}d\theta^{\prime}\left(\ln\left|\epsilon e^{i\theta}-\epsilon e^{i\theta^{\prime}}\right|+\ln\left|\epsilon e^{i\theta}-\epsilon e^{-i\theta^{\prime}}\right|\right)\right]\right\} \,,\label{eq:primary}
\end{eqnarray}
where $\mathrm{R}\left[\cdots\right]$ denotes the radial ordering.
Substituting the mode expansions (\ref{eq:mode_a}), (\ref{eq:mode})
into (\ref{eq:primary}), it is straightforward to calculate the expectation
value on the right-hand side of (\ref{eq:primary}) and we obtain
\begin{eqnarray}
 &  & \left\langle \ \right|\mathrm{R}\exp\left[\frac{i}{\pi}\int_{0}^{\pi}d\theta\left(p\tilde{X}(z+\epsilon e^{i\theta},\bar{z}+\epsilon e^{-i\theta})+\tilde{p}X(z+\epsilon e^{i\theta},\bar{z}+\epsilon e^{-i\theta})\right)\right]\left|\ \right\rangle ^{\prime}\nonumber \\
 &  & \quad=\exp\left[-\frac{\alpha^{\prime}p\tilde{p}}{4\pi^{2}}\int_{0}^{\pi}d\theta\int_{0}^{\pi}d\theta^{\prime}\left(\frac{1}{\lambda}\left(\frac{w^{\prime}}{w}\right)^{\lambda}F(\lambda,1,1+\lambda;\frac{w^{\prime}}{w})+\frac{1}{\lambda}\left(\frac{\bar{w}^{\prime}}{w}\right)^{\lambda}F(\lambda,1,1+\lambda;\frac{\bar{w}^{\prime}}{w})\right.\right.\nonumber \\
 &  & \hphantom{\quad=\exp-\frac{\alpha^{\prime}p\tilde{p}}{4\pi^{2}}\int_{0}^{\pi}d\theta\int_{0}^{\pi}d\theta^{\prime}\quad}+\frac{1}{\lambda}\left(\frac{w^{\prime}}{\bar{w}}\right)^{\lambda}F(\lambda,1,1+\lambda;\frac{w^{\prime}}{\bar{w}})+\frac{1}{\lambda}\left(\frac{\bar{w}^{\prime}}{\bar{w}}\right)^{\lambda}F(\lambda,1,1+\lambda;\frac{\bar{w}^{\prime}}{\bar{w}})\nonumber \\
 &  & \hphantom{\quad=\exp-\frac{\alpha^{\prime}p\tilde{p}}{4\pi^{2}}\int_{0}^{\pi}d\theta\int_{0}^{\pi}d\theta^{\prime}\quad}+\frac{1}{1-\lambda}\left(\frac{w}{w^{\prime}}\right)^{1-\lambda}F(1-\lambda,1,2-\lambda;\frac{w}{w^{\prime}})\nonumber \\
 &  & \hphantom{\quad=\exp-\frac{\alpha^{\prime}p\tilde{p}}{4\pi^{2}}\int_{0}^{\pi}d\theta\int_{0}^{\pi}d\theta^{\prime}\quad}+\frac{1}{1-\lambda}\left(\frac{\bar{w}}{w^{\prime}}\right)^{1-\lambda}F(1-\lambda,1,2-\lambda;\frac{\bar{w}}{w^{\prime}})\nonumber \\
 &  & \hphantom{\quad=\exp-\frac{\alpha^{\prime}p\tilde{p}}{4\pi^{2}}\int_{0}^{\pi}d\theta\int_{0}^{\pi}d\theta^{\prime}\quad}+\frac{1}{1-\lambda}\left(\frac{w}{\bar{w}^{\prime}}\right)^{1-\lambda}F(1-\lambda,1,2-\lambda;\frac{w}{\bar{w}^{\prime}})\nonumber \\
 &  & \hphantom{\quad=\exp-\frac{\alpha^{\prime}p\tilde{p}}{4\pi^{2}}\int_{0}^{\pi}d\theta\int_{0}^{\pi}d\theta^{\prime}\quad}\left.\left.+\frac{1}{1-\lambda}\left(\frac{\bar{w}}{\bar{w}^{\prime}}\right)^{1-\lambda}F(1-\lambda,1,2-\lambda;\frac{\bar{w}}{\bar{w}^{\prime}})\right)\right]\nonumber \\
 &  & \hphantom{\quad=\quad}\times\left\langle \ \right|e^{i(p\tilde{x}+\tilde{p}x)}\left|\ \right\rangle ^{\prime}\,,\label{eq:langleRexp}
\end{eqnarray}
where $F(\alpha,\beta,\gamma;\,z)$ is the hypergeometric function\footnote{We note the relation: 
\[
\frac{1}{\lambda}F(\lambda,1,1+\lambda;z)=\frac{\pi}{\sin\pi\lambda}(-z)^{-\lambda}+\frac{1}{(1-\lambda)z}\,F(1-\lambda,1,2-\lambda;\frac{1}{z})\ \ \ (\,|{\rm arg}(-z)|<\pi\,).
\]
} and 
\begin{eqnarray*}
w & = & z+\epsilon e^{i\theta}\,,\\
w^{\prime} & = & z+\epsilon e^{i\theta^{\prime}}\,.
\end{eqnarray*}
Using the formula \cite{abramowitz} 
\begin{align*}
F(\alpha,\beta,\alpha+\beta;\,z) & =\frac{\Gamma(\alpha+\beta)}{\Gamma(\alpha)\Gamma(\beta)}\sum_{n=0}^{\infty}\frac{(\alpha)_{n}(\beta)_{n}}{(n!)^{2}}\left\{ 2\psi(n+1)-\psi(n+\alpha)-\psi(n+\beta)-\ln(1-z)\right\} (1-z)^{n}\\
 & =\frac{\Gamma(\alpha+\beta)}{\Gamma(\alpha)\Gamma(\beta)}\left\{ -\ln(1-z)+2\psi(1)-\psi(\alpha)-\psi(\beta)\right\} +\cdots
\end{align*}
for $\left|\mathrm{arg}(1-z)\right|<\pi,\,\left|1-z\right|<1$, the
right-hand side of (\ref{eq:primary}) is evaluated to be 
\[
{\displaystyle \frac{\delta^{-\frac{\alpha'p^{2}}{2}}}{|z|^{\alpha'p^{2}}}}\left\langle \ \right|e^{i(p\tilde{x}+\tilde{p}x)}\left|\ \right\rangle ^{\prime}\,,
\]
where 
\begin{equation}
\ln\delta=2\psi(1)-\psi(\lambda)-\psi(1-\lambda)\,,\label{eq:delta}
\end{equation}
and $\psi(x)$ is the digamma function.

For $z<0$, the three-point function (\ref{eq:three-point}) can be
calculated in the same way using 
\begin{eqnarray}
 &  & \left\langle \ \right|e^{i\vec{p}\cdot\vec{X}}(z,\bar{z})\left|\ \right\rangle ^{\prime}\nonumber \\
 &  & \quad=\lim_{\epsilon\to0}\left\{ \left\langle \ \right|\mathrm{R}\exp\left[\frac{i}{\pi}\int_{0}^{\pi}d\theta\left(p\tilde{X}(z+\epsilon e^{i\theta},\bar{z}+\epsilon e^{-i\theta})+\tilde{p}X(z+\epsilon e^{i\theta},\bar{z}+\epsilon e^{-i\theta})\right)\right]\left|\ \right\rangle ^{\prime}\vphantom{\left[-\frac{\alpha^{\prime}p\tilde{p}}{\pi^{2}}\int_{0}^{\pi}d\theta\int_{0}^{\pi}d\theta^{\prime}\left(\ln\left|\epsilon e^{i\theta}-\epsilon e^{i\theta^{\prime}}\right|+\cos2\pi\lambda\ln\left|\epsilon e^{i\theta}-\epsilon e^{-i\theta^{\prime}}\right|\right)\right]}\right.\nonumber \\
 &  & \quad\hphantom{=\lim_{\epsilon\to0}\exp}\times\left.\exp\left[-\frac{\alpha^{\prime}p\tilde{p}}{\pi^{2}}\int_{0}^{\pi}d\theta\int_{0}^{\pi}d\theta^{\prime}\left(\ln\left|\epsilon e^{i\theta}-\epsilon e^{i\theta^{\prime}}\right|+(\cos2\pi\lambda)\ln\left|\epsilon e^{i\theta}-\epsilon e^{-i\theta^{\prime}}\right|\right)\right]\right\}\nonumber \\
 &  & \ \label{eq:primary-1}
\end{eqnarray}
and (\ref{eq:langleRexp}) with $z=e^{\pi i}|z|$. We eventually obtain
\[
\left\langle \ \right|e^{i\vec{p}\cdot\vec{X}}(z,\bar{z})\left|\ \right\rangle ^{\prime}=\begin{cases}
{\displaystyle \frac{\delta^{-\frac{\alpha'p^{2}}{2}}}{|z|^{\alpha'p^{2}}}}\left\langle \ \right|e^{i(p\tilde{x}+\tilde{p}x)}\left|\ \right\rangle ^{\prime} & (z>0)\\
\\
{\displaystyle \frac{\delta^{-\frac{\alpha'p^{2}}{2}\cos^{2}\pi\lambda}}{|z|^{\alpha'p^{2}\cos^{2}\pi\lambda}}}\left\langle \ \right|e^{i(p\tilde{x}+\tilde{p}x)}\left|\ \right\rangle ^{\prime} & (z<0)\,.
\end{cases}
\]

\section{Correlation functions of four BCC operators\label{sec:Green's-function-with}}

The four-point functions of the BCC operators 
\[
\left<\bar{\sigma}_{*}^{j}(x_{1})\sigma_{*}^{i}(x_{2})\bar{\sigma}_{*}^{k}(x_{3})\sigma_{*}^{l}(x_{4})\right>
\]
can be calculated using the the technique developed in the orbifold
CFT \cite{Dixon1987,Gava1997b}. The correlation function can be given
as 
\[
\left<\bar{\sigma}_{*}^{j}(x_{1})\sigma_{*}^{i}(x_{2})\bar{\sigma}_{*}^{k}(x_{3})\sigma_{*}^{l}(x_{4})\right>=\sum_{\vec{p}}C_{\vec{p}}\left<\bar{\sigma}_{*}^{j}(x_{1})\sigma_{*}^{i}(x_{2})\bar{\sigma}_{*}^{k}(x_{3})\sigma_{*}^{l}(x_{4})\right>_{\vec{p}}\,,
\]
where $\left<\bar{\sigma}_{*}^{j}(x_{1})\sigma_{*}^{i}(x_{2})\bar{\sigma}_{*}^{k}(x_{3})\sigma_{*}^{l}(x_{4})\right>_{\vec{p}}$
is the conformal block specified by the momenta $\vec{p}$ of the
operators which appear in the OPE of $\bar{\sigma}_{*}^{j}(x_{1})$
and $\sigma_{*}^{i}(x_{2})$.

Let us consider the following ratio of correlation functions defined
on the upper half plane with the insertions of the BCC operators at
$x_{1}<x_{2}<x_{3}<x_{4}$: 
\begin{equation}
\tilde{G}_{\vec{p}}(z,\,w,\,x_{i})=\frac{\left<\partial X(z)\partial\tilde{X}(w)\bar{\sigma}_{*}^{j}(x_{1})\sigma_{*}^{i}(x_{2})\bar{\sigma}_{*}^{k}(x_{3})\sigma_{*}^{l}(x_{4})\right>_{\vec{p}}}{\left<\bar{\sigma}_{*}^{j}(x_{1})\sigma_{*}^{i}(x_{2})\bar{\sigma}_{*}^{k}(x_{3})\sigma_{*}^{l}(x_{4})\right>_{\vec{p}}}\,.\label{eq:Greenfunc-1}
\end{equation}
As a function of $z,w$, $\tilde{G}_{\vec{p}}(z,\,w,\,x_{i})$ can
be analytically continued to be defined on the whole complex plane,
which corresponds to the double of the worldsheet. Since $\tilde{G}_{\vec{p}}(z,\,w,\,x_{i})$
is a 1-form on the sphere with respect to variables $z,w$, it behaves
like $\sim{\cal O}(z^{-2})\ (z\sim\infty)$ and $\sim{\cal O}(w^{-2})\ (w\sim\infty)$.
From the OPE of $\partial X$ and $\partial\tilde{X}$, it follows
that 
\begin{equation}
\tilde{G}_{\vec{p}}(z,\,w,\,x_{i})=\frac{-\alpha'/2}{(z-w)^{2}}+{\rm finite}\ \ \ \ (z\sim w)\,.\label{eq:Gzw}
\end{equation}
From the OPE between $\partial X,\,\partial\tilde{X}$ and the BCC operators,
we get 
\begin{equation}
\tilde{G}_{\vec{p}}(z,\,w,\,x_{i})\sim\begin{cases}
(z-x_{i})^{-(1-\lambda)} & (z\sim x_{i},\ i=1,3)\\
(z-x_{i})^{-\lambda} & (z\sim x_{i},\ i=2,4)\\
(w-x_{i})^{-\lambda} & (w\sim x_{i},\ i=1,3)\\
(w-x_{i})^{-(1-\lambda)} & (w\sim x_{i},\ i=2,4)\,.
\end{cases}\label{eq:Gzwx}
\end{equation}
Using all these conditions, $\tilde{G}_{\vec{p}}(z,\,w,\,x_{i})$
is fixed as 
\begin{eqnarray*}
\tilde{G}_{\vec{p}}(z,\,w;\,x_{i}) & = & -\frac{\alpha'}{2}\omega(z)\tilde{\omega}(w)\left[(1-\lambda)\frac{(z-x_{1})(z-x_{3})(w-x_{2})(w-x_{4})}{(z-w)^{2}}\right.\\
 &  & \ \ \ \left.+\lambda\frac{(z-x_{2})(z-x_{4})(w-x_{1})(w-x_{3})}{(z-w)^{2}}+A(x_{i})\right]\,,
\end{eqnarray*}
where the functions $\omega(z)$ and $\tilde{\omega}(w)$ are given
by 
\begin{align}
\omega(z) & =(z-x_{1})^{-1+\lambda}(z-x_{2})^{-\lambda}(z-x_{3})^{-1+\lambda}(z-x_{4})^{-\lambda}\,,\\
\tilde{\omega}(w) & =(w-x_{1})^{-\lambda}(w-x_{2})^{-1+\lambda}(w-x_{3})^{-\lambda}(w-x_{4})^{-1+\lambda}\,,
\end{align}
and $A(x_{i})$ is a meromorphic function which is left undetermined.
We can use $SL(2,R)$ transformation to fix $x_{1}=0$, $x_{2}=x$,
$x_{3}=1$, and $x_{4}\rightarrow\infty$, and $\tilde{G}(z,\,w;\,x)$
becomes 
\begin{equation}
\tilde{G}_{\vec{p}}(z,\,w;\,x)=-\frac{\alpha'}{2}\omega(z)\tilde{\omega}(w)\left[(1-\lambda)\frac{z(z-1)(w-x)}{(z-w)^{2}}+\lambda\frac{(z-x)w(w-1)}{(z-w)^{2}}+A(x)\right]\,,\label{eq:GreenfuncA}
\end{equation}
where $\omega(z)=z^{-1+\lambda}(z-x)^{-\lambda}(z-1)^{-1+\lambda}$
and $\tilde{\omega}(w)=w^{-\lambda}(w-x)^{-1+\lambda}(w-1)^{-\lambda}$.

$A(x)$ can be fixed by using the conditions 
\begin{equation}
\oint dz\,\partial X=2\pi\alpha'p,\ \ \ \oint dz\,\partial\tilde{X}=2\pi\alpha'\tilde{p}\label{eq:int_dX}
\end{equation}
where the integration contours are around the branch cut on the interval
($0,\,x)$. $p$ and $\tilde{p}$ are the momenta of the operators
which appear in the OPE of $\bar{\sigma}_{*}^{j}(0)$ and $\sigma_{*}^{i}(x)$.
Integrating (\ref{eq:GreenfuncA}) around the branch cut in the interval
$(0,\,x)$ and using the condition (\ref{eq:int_dX}), we obtain 
\begin{equation}
\oint dz\,\tilde{G}_{\vec{p}}(z,\,w,\,x)=2ie^{\pi i\lambda}\sin\pi\lambda\int_{0}^{x}dz\,\tilde{G}_{\vec{p}}(z,\,w,\,x)=2\pi\alpha'p\frac{\left<\partial\tilde{X}(w)\bar{\sigma}_{*}^{j}(0)\sigma_{*}^{i}(x)\bar{\sigma}_{*}^{k}(1)\sigma_{*}^{l}(\infty)\right>_{\vec{p}}}{\left<\bar{\sigma}_{*}^{j}(0)\sigma_{*}^{i}(x)\bar{\sigma}_{*}^{k}(1)\sigma_{*}^{l}(\infty)\right>_{\vec{p}}}\,.\label{eq:intGtilde}
\end{equation}
The quantity on the rightmost side of (\ref{eq:intGtilde}) can be
seen to be proportional to $\tilde{\omega}(w)$ and the proportionality
factor can be determined by imposing (\ref{eq:int_dX}) once more.
As a result, we get 
\begin{equation}
\int_{0}^{x}dz\,\tilde{G}_{\vec{p}}(z,\,w;\,x)=\frac{\pi^{2}\alpha'^{2}p^{2}}{2\sin^{2}\pi\lambda}\frac{\tilde{\omega}(w)}{\int_{0}^{x}dw\,\tilde{\omega}(w)}\,.
\end{equation}
Taking the limit $w\rightarrow\infty$, we find that $A(x)$ is given
by 
\begin{equation}
A(x)=-\lambda\frac{\int_{0}^{x}dz\,(z-x)\omega(z)}{\int_{0}^{x}dz\,\omega(z)}-\frac{\pi^{2}\alpha'p^{2}}{\sin^{2}\pi\lambda}\frac{1}{\int_{0}^{x}dz\,\omega(z)\int_{0}^{x}dw\,\tilde{\omega}(w)}\,.
\end{equation}
All the integrals which appear on the right-hand side can be expressed
in terms of the hypergeometric function: 
\begin{align}
\int_{0}^{x}dz\,\omega(z) & =-\frac{\pi}{\sin\pi\lambda}F(\lambda,1-\lambda,1;x)\,,\\
\int_{0}^{x}dw\,\tilde{\omega}(w) & =-\frac{\pi}{\sin\pi\lambda}F(\lambda,1-\lambda,1;x)\,,\\
\int_{0}^{x}dz\,(z-x)\omega(z) & =\frac{\pi}{\lambda\sin\pi\lambda}x(1-x)\frac{d}{dx}F(\lambda,1-\lambda,1;x)\,,\\
\frac{1}{\int_{0}^{x}dz\,\omega(z)\int_{0}^{x}dw\,\tilde{\omega}(w)} & =-\frac{\sin\pi\lambda}{\pi}x(1-x)\frac{d}{dx}\frac{F(\lambda,1-\lambda,1;1-x)}{F(\lambda,1-\lambda,1;x)}\,.
\end{align}
Here we have used the following formula for the hypergeometric function
\cite{abramowitz}, 
\begin{align}
F(\alpha,\beta,\gamma;\,z) & =\frac{\Gamma(\gamma)}{\Gamma(\beta)\Gamma(\gamma-\beta)}\int_{0}^{1}t^{\beta-1}(1-t)^{\gamma-\beta-1}(1-tz)^{-\alpha}\,dt\,,\\
\frac{d}{dz}\left[(1-z)^{\alpha+\beta-\gamma}F(\alpha,\beta,\gamma;\,z)\right] & =\frac{(\gamma-\alpha)(\gamma-\beta)}{\gamma}(1-z)^{\alpha+\beta-\gamma-1}F(\alpha,\beta,\gamma+1;\,z)\,,\\
\frac{d}{dz}[z^{\gamma-1}F(\alpha,\beta,\gamma;z)] & =(\gamma-1)z^{\gamma-2}F(\alpha,\beta,\gamma-1;z)\,,\\
\frac{d}{dz}\left[\frac{F(\alpha,1-\alpha,1;\,1-z)}{F(\alpha,1-\alpha,1;\,z)}\right] & =-\frac{\sin\pi\alpha}{\pi}\frac{z^{-1}(1-z)^{-1}}{F(\alpha,1-\alpha,1;\,z)^{2}}\,.
\end{align}

Thus we eventually find 
\begin{align}
\tilde{G}_{\vec{p}}(z,\,w;\,x) & =-\frac{\alpha'}{2}\omega(z)\tilde{\omega}(w)\left[(1-\lambda)\frac{z(z-1)(w-x)}{(z-w)^{2}}+\lambda\frac{(z-x)w(w-1)}{(z-w)^{2}}\right.\nonumber \\
 & \ \ \ \left.+x(1-x)\frac{d}{dx}\ln F(\lambda,1-\lambda,1;x)+\frac{\pi\alpha'p^{2}}{\sin\pi\lambda}\,x(1-x)\frac{d}{dx}\frac{F(\lambda,1-\lambda,1;1-x)}{F(\lambda,1-\lambda,1;x)}\right]\,.\label{eq:GreenfuncF}
\end{align}
From this correlation function, we can derive 
\begin{eqnarray*}
 &  & \frac{\left<T(z)\bar{\sigma}_{*}^{j}(0)\sigma_{*}^{i}(x)\bar{\sigma}_{*}^{k}(1)\sigma_{*}^{l}(\infty)\right>_{\vec{p}}}{\left<\bar{\sigma}_{*}^{j}(0)\sigma_{*}^{i}(x)\bar{\sigma}_{*}^{k}(1)\sigma_{*}^{l}(\infty)\right>_{\vec{p}}}\\
 &  & \quad=\frac{1}{2}\lambda(1-\lambda)\left(\frac{1}{z}+\frac{1}{z-1}-\frac{1}{z-x}\right)^{2}\\
 &  & \hphantom{\quad=\frac{1}{2}}+\frac{x(1-x)}{z(z-1)(z-x)}\frac{d}{dx}\left[\ln F(\lambda,1-\lambda,1;x)+\frac{\pi\alpha'p^{2}}{\sin\pi\lambda}\frac{F(\lambda,1-\lambda,1;1-x)}{F(\lambda,1-\lambda,1;x)}\right]
\end{eqnarray*}
using (\ref{eq:EM}). Considering the limit $z\to x$, we get 
\begin{eqnarray*}
 &  & \frac{d}{dx}\ln\left<\bar{\sigma}_{*}^{j}(0)\sigma_{*}^{i}(x)\bar{\sigma}_{*}^{k}(1)\sigma_{*}^{l}(\infty)\right>_{\vec{p}}\\
 &  & \quad=-\lambda(1-\lambda)\left(\frac{1}{x}+\frac{1}{x-1}\right)\\
 &  & \hphantom{\quad=-\lambda}-\frac{d}{dx}\left[\ln F(\lambda,1-\lambda,1;x)+\frac{\pi\alpha'p^{2}}{\sin\pi\lambda}\frac{F(\lambda,1-\lambda,1;1-x)}{F(\lambda,1-\lambda,1;x)}\right]\,.
\end{eqnarray*}
Therefore we find that the four-point function can be expressed as
\begin{eqnarray}
 &  & \left<\bar{\sigma}_{*}^{j}(0)\sigma_{*}^{i}(x)\bar{\sigma}_{*}^{k}(1)\sigma_{*}^{l}(\infty)\right>\nonumber \\
 &  & \quad=\sum_{\vec{p}}C_{\vec{p}}\,x^{-\lambda(1-\lambda)}(1-x)^{-\lambda(1-\lambda)}\frac{1}{F(\lambda,1-\lambda,1;x)}\exp\left[-\frac{\pi\alpha'}{\sin\pi\lambda}p^{2}\frac{F(\lambda,1-\lambda,1;1-x)}{F(\lambda,1-\lambda,1;x)}\right]\,,\nonumber \\
 &  & \ \label{eq:4ptfunc_p}
\end{eqnarray}
where the constant $C_{\vec{p}}$ is determined by taking the limit
$x\rightarrow0$ in (\ref{eq:4ptfunc_p}). Taking the limit of the
right-hand side of (\ref{eq:4ptfunc_p}), we get 
\begin{equation}
\left<\bar{\sigma}_{*}^{j}(0)\sigma_{*}^{i}(x)\bar{\sigma}_{*}^{k}(1)\sigma_{*}^{l}(\infty)\right>\sim\sum_{\vec{p}}C_{p}\delta^{-\alpha'p^{2}}x^{-\lambda(1-\lambda)+\alpha'p^{2}}\,,\label{eq:4ptfunc-limit2}
\end{equation}
where $\delta$ is the one which appears in (\ref{eq:delta}). On
the other hand, using the OPE (\ref{eq:OPE*}), we obtain 
\begin{eqnarray}
 &  & \left<\bar{\sigma}_{*}^{j}(0)\sigma_{*}^{i}(x)\bar{\sigma}_{*}^{k}(1)\sigma_{*}^{l}(\infty)\right>\nonumber \\
 &  & \quad\sim\frac{1}{\left(2\pi\right)^{2}R_{1}R_{2}}x^{-\lambda(1-\lambda)}\sum_{n_{1},n_{2}}(x\delta^{-\frac{1}{2}})^{\alpha^{\prime}\left(\left(\frac{n_{1}}{R_{1}}\right)^{2}+\left(\frac{n_{2}}{R_{2}}\right)^{2}\right)}\omega^{\frac{n_{1}n_{2}}{2}+n_{2}j}\delta_{i-j,\,n_{1}\,({\rm mod}N)}\left<\bar{\sigma}_{*}^{k}\right|e^{-ip\cdot X(1,1)}\left|\sigma_{*}^{l}\right>\,,\nonumber \\
 &  & \ \label{eq:4ptfunc-limit1}
\end{eqnarray}
where $p_{i}=n_{i}/R_{i}$($i=1,2$). Substituting (\ref{eq:3ptfunc-torus})
into (\ref{eq:4ptfunc-limit1}) and comparing it with (\ref{eq:4ptfunc-limit2}),
we can derive 
\begin{equation}
C_{p}=\frac{1}{\left(2\pi\right)^{2}R_{1}R_{2}}\omega^{n_{1}n_{2}+n_{2}(j-k)}\delta_{i-j,n_{1}\,({\rm mod}N)}\delta_{l-k,-n_{1}\,({\rm mod}N)}
\end{equation}
and 
\begin{eqnarray}
 &  & \left<\bar{\sigma}_{*}^{j}(0)\sigma_{*}^{i}(x)\bar{\sigma}_{*}^{k}(1)\sigma_{*}^{l}(\infty)\right>\nonumber \\
 &  & \quad=\frac{1}{\left(2\pi\right)^{2}R_{1}R_{2}}x^{-\lambda(1-\lambda)}(1-x)^{-\lambda(1-\lambda)}\frac{1}{F(\lambda,1-\lambda,1;x)}\,\delta_{j-i,l-k\,{\rm (mod}N)}\nonumber \\
 &  & \hphantom{\quad=}\times\sum_{n,m}\omega^{(j-l)m}\exp\left[-\frac{\pi\alpha'}{\sin\pi\lambda}\frac{F(\lambda,1-\lambda,1;1-x)}{F(\lambda,1-\lambda,1;x)}\left\{ \frac{(j-i+nN)^{2}}{R_{1}^{2}}+\frac{m^{2}}{R_{2}^{2}}\right\} \right]\,.\label{eq:four-point}
\end{eqnarray}
Here, the sums over $n,m$ correspond to those over the momenta in the
$x^{1}$ and $x^{2}$ directions. The momenta in these directions
are treated asymmetrically because of the choice of the states corresponding
to the BCC operators.

The noncompact case can be dealt with in the same way. The sum over
the intermediate momenta in (\ref{eq:4ptfunc_p}) becomes an integral
over them and we get (\ref{eq:4ptfunc-infty}).

 \bibliographystyle{utphys}
\bibliography{reference}

\end{document}